\documentclass{ucetd}
\usepackage{subfigure,epsfig,amsfonts}
\usepackage{natbib}
\usepackage{amsmath}
\usepackage{amssymb}
\usepackage{amsthm}
\setcitestyle{square,numbers}
\newcommand{\be}{\begin{equation}}	
\newcommand{\ee}{\end{equation}}
\newcommand{\tr}{\mbox{tr}}
\newcommand{\eg}{{\it e.g.}}
\newcommand{\lsb}{\left[}
\newcommand{\rsb}{\right]}
\newcommand{\lb}{\left(}
\newcommand{\rb}{\right)}
\newcommand{\EE}{\mathcal{E}}

\title{Bulk Locality from Entanglement in Gauge/Gravity Duality} 
\author{Jennifer Lin}
\department{Physics}
\division{Physical Sciences} 
\degree{Doctor  of Philosophy} 
\date{August 2015}

\begin{document}
\maketitle


\tableofcontents
\listoffigures

\acknowledgments
First and foremost, I am very deeply grateful to my advisor David Kutasov for his guidance over the past five years. 
I learned so much from our conversations.

I've benefitted a lot from interactions with many people in the physics department and theory group at Chicago. I would especially like to thank Jeff Harvey, Cheng Chin and Michael Levin for serving on my thesis committee. 

This thesis is based on a project with Matilde Marcolli, Hirosi Ooguri and Bogdan Stoica. I am grateful to them for the collaboration as well as continued discussion of related issues, and to Hirosi and Caltech for hosting me in April 2014. Thanks to David for comments on an early draft of the first chapter. 
I'd also like to thank all my collaborators on other projects (and attempted projects!) for teaching me lots of interesting things.
		
		I am indebted to several people at Princeton, where I first got interested in this field. I am especially grateful to Lyman Page for introducing me to research and to Herman Verlinde for introducing me to string theory. Igor Klebanov, Silviu Pufu and Paul Steinhardt also greatly influenced my early development as a physicist.	
		
	Special thanks to Djordje Radicevic and Bogdan Stoica for all the chats ever since we were undergrads going over problem sets together. Thanks also to Mike Geracie and Yiwei She for fun discussions about physics and  friendship outside of it.

Finally, thanks to my sister Stephanie who has always been a best friend, to my parents for their encouragement all my life, and to Jae for showing me how to write stories.
  
My work was supported by a NSF Graduate Research Fellowship and by the Bloomenthal Fellowship from the University of Chicago.

\abstract
Gauge/gravity duality posits an equivalence between certain strongly coupled quantum field theories and theories of gravity with negative cosmological constant in a higher number of spacetime dimensions. The map between the degrees of freedom on the two sides is non-local and incompletely understood. I describe recent work towards characterizing this map using entanglement in the QFT, where near the dual AdS boundary, the classical energy density at a point in the bulk is stored in the relative entropies of boundary subregions whose homologous minimal surfaces pass through the bulk point. I also derive bulk classical energy conditions  near the AdS boundary from entanglement inequalities in the CFT. This is based on the paper \cite{Lin:2014hva} with Matilde Marcolli, Hirosi Ooguri and Bogdan Stoica. 

More generally, in recent years, there has appeared some evidence that quantum entanglement is responsible for the emergence of spacetime. I review and comment on the state of these developments.
\\
\\
Note added for arXiv version: Section 1.5 (especially 1.5.2, 1.5.6, 1.5.7), with 1.2.2 as intro/motivation, contain the bulk of comments pertinent to ``entanglement and spacetime". Chapter 2 overlaps with [1]. There are no new quantitative results. 

\mainmatter

\chapter{Introduction}

\section{Overview}

One of the great challenges for fundamental physics is to formulate a quantum theory of gravity. According to the classical theory, general relativity (GR), spacetime is 
	dynamical; it is warped by matter, even as it tells matter to move along geodesics, which reduces to the earlier Newtonian concept of gravity. 
%
%
More precisely, general relativity is a classical field theory of the metric that encodes distances in and thereby contains all local information about spacetime viewed as a manifold.
Therefore, a quantum theory of gravity is expected to shed light on some of the deepest questions for science such as the structure of space at the smallest scales, the nature of time, and the nature of the Big Bang. 

But GR has proved resistant to quantization by the methods of quantum field theory (QFT), the theory that describes the other fundamental forces to unprecedented accuracy. The essence of the conflict is that while QFT contains local operators defined at each point of a fixed underlying spacetime, such objects are forbidden in the framework of GR, where they are gauge-variant under  diffeomorphism symmetry.
%
%

String theory, which replaces pointlike degrees of freedom with extended objects, 
is the only known consistent quantum framework that contains gravity. The theory was originally formulated in the 1960's to describe the strong nuclear force, but had an embarrassing flaw for this purpose: its spectrum  contained a massless spin-2 particle. In the 1970's, the spin-2 mode was reinterpreted as the graviton, and the theory took on its modern guise as a candidate theory of quantum gravity. 

 Prior to the mid-1990's, string theory was mostly studied perturbatively around fixed backgrounds. In this approach, one studies the theory on the worldsheet that a string sweeps out in spacetime, which is a 2d theory of quantum gravity with spacetime coordinates appearing as dynamical fields, and the background metric as a K\"ahler potential for them. 
 (2d gravity is much simpler than its higher-dimensional cousins because one can locally gauge-fix all the metric degrees of freedom.)
String S-matrix elements are computed by evaluating correlators of vertex operators on the worldsheet.
The perturbative expansion in the string coupling constant $g_s$
 requires summing over all topologies for the worldsheet CFT to live on, so
in practical terms, this approach is mostly good for calculating S-matrix elements to the leading order or first order correction in $g_s$.
But from looking at the worldsheet theory itself, people discovered some interesting features. Comparisons of different but equivalent descriptions of the worldsheet showed that there are situations where apparently different spacetimes give rise to the same physics to all orders in string perturbation theory. Only one background at a time is large in string units, but both backgrounds are equally good at the string scale. The simplest example, T-duality, relates strings propagating on a spacetime circle of radius $R$ to one of radius $\ell_s^2/R$. Mirror symmetry which is its application to fibers of a Calabi-Yau, 
  shows that even the topology of the background need not be invariant under  different ways of viewing the theory.
  (See \cite{Martinec:2013hsa} for a nice summary of these developments). 

%

In the mid-1990's there were two major and not unrelated developments: the discovery of many S-dualities in QFT and string theory (discussed more in section 1.2.1) and the discovery and extensive study of branes in various limits. D-branes are solitons in string theory, on which open strings can end. Their worldvolume dynamics is characterized by an open string theory, which at energies well below the string scale, reduces to an ordinary QFT. A consequence is that branes nicely geometrize many features of QFT's, see \cite{Giveon:1998sr} for a review. E.g.
%
%
 many QFT S-dualities were shown to arise simply by moving systems of intersecting branes around and taking the low-energy worldvolume limit. 

On the other hand, as energetic objects in a theory containing gravity, the branes backreact on spacetime. In particular, large numbers of D-branes at large $g_s$ source ten-dimensional supergravity (SUGRA) generalizations of black hole backgrounds. 

It was soon realized that much could be gained from interpolating between the worldvolume and supergravity descriptions. One early such construction by Strominger and Vafa \cite{Strominger:1996sh} successfully accounted for black hole microstates in string theory (by comparing the entropy of the worldvolume CFT on a system of branes to the horizon area in the supergravity limit, where supersymmetry protected the number of states as one dialed the coupling.) But the most celebrated example of this paradigm is the one that I review next.
  
   \section{Gauge/gravity duality}

Consider a stack of $N$ D3 branes in Type IIB string theory. The branes backreact on spacetime with a strength of $G_{10}^N\cdot N \cdot T_3 = g_s N \alpha'^2$, where $G_{10}^N \sim g_s^2\alpha'^4$ is the ten-dimensional Newton constant (that can be read off comparing the 3-graviton tree graph with the equivalent diagram in string theory) 
and $T_3 \sim g_s^{-1}\alpha'^{-2}$ is the tension of a single D3 brane. 
Thus, at small $\lambda = g_s \cdot N$ in string units, their effect on the geometry of a ${\bf R}^{9,1}$  background  is negligible, but at large $\lambda$, they curve the transverse geometry to a black three-brane SUGRA background. In other words the first step is to identify the D-branes, viewed as endpoints for open strings, with p-brane SUGRA solutions. Now consider string theory in the presence of the branes. In the first case there are two interacting sectors, the four-dimensional open string theory on the D3 brane worldvolume along with closed strings in the ten dimensional bulk. Below the string scale, the worldvolume theory becomes 4d $N=4$ supersymmetric Yang-Mills theory (SYM) with gauge group $SU(N)$ (after one removes the mode describing center-of-mass motion of the branes). In the second case, string theory around the p-brane background consists of closed strings propagating both near and far away from the black brane horizon. 

The key idea of Maldacena in 1997 was that one can take a certain limit of the D3 brane system, whose action is to decouple the open string sector from the closed string one at small $\lambda$, and zoom in on the near-horizon limit of the SUGRA background at large $\lambda$, which turns out to be $AdS_5 \times {\bf S}^5$ (with RR flux). Observing two descriptions of the same system in different parts of parameter space, he conjectured \cite{Maldacena:1997re}: 

\begin{quote}
{\it Type IIB string theory in asymptotically $AdS_5\times {\bf S}^5$ spacetime is exactly equivalent to 4d N=4 SU(N) SYM, a local QFT, }
\end{quote}
with identifications of dimensionless parameters that is reviewed below.
The meaning of `exact equivalence' is that the theories are really different descriptions of the same physical system: Any physical question asked of one theory can in principle be formulated and answered in the other (though after 18 years, the exact nature of the map is still far from being fully elucidated).
 Called the AdS/CFT duality, this provides an explicit non-perturbative, background-independent definition of string theory and realizes the holographic equivalence of quantum gravity to a QFT in one lower noncompact dimension.

\begin{figure}[h]
\centering
\includegraphics[width=0.65\textwidth]{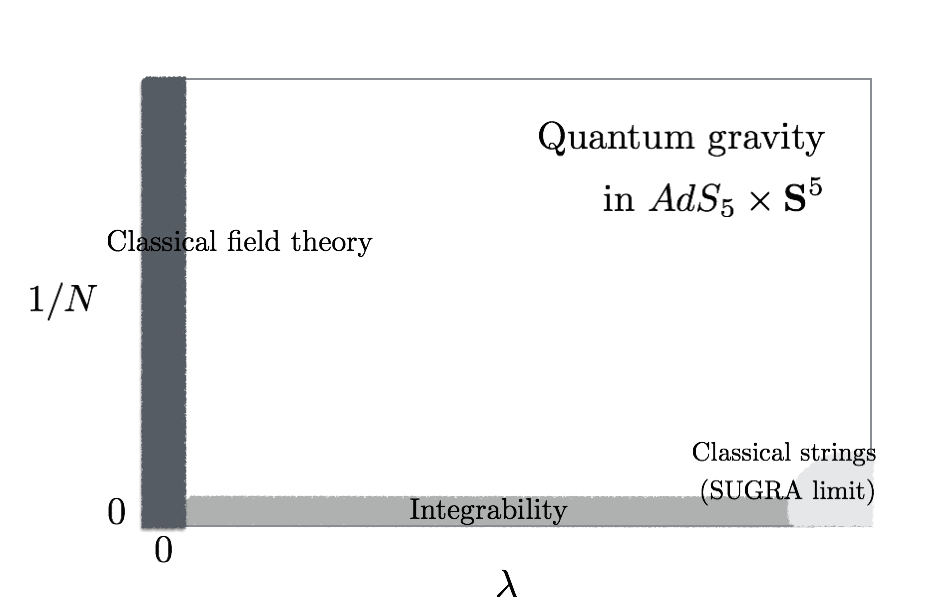} 
    \caption{The parameter space of $N=4$ $SU(N)$ SYM with 't Hooft coupling $\lambda$. 
    Over each point in this picture there is a Hilbert space that can be interpreted as either the Hilbert space of the CFT, or that of string theory in asymptotically AdS$_5 \times {\bf S}^5$ spacetime with the identification of dimensionless parameters $\ell_{AdS}/\ell_s = \lambda^{1/4}$ and $\ell_{AdS}/\ell_P \sim N^{1/4}$. The shaded regions depict ranges of the parameters for which we can do some computations. This figure is adapted from [13].}
\end{figure}

It is straightforward to repeat Maldacena's argument for other systems of branes, giving other top-down examples of gauge/gravity dualities. In fact, the modern version of the conjecture \cite{Heemskerk:2009pn} goes well beyond such   considerations: \footnote{There are examples of holographic duality where the dual to a CFT seems to be at an intermediate level of complexity between classical gravity and full-blown string theory -- i.e. higher-spin gravity \cite{Klebanov:2002ja, Douglas:2010rc, Giombi:2012ms} -- but  at least in some cases, higher-spin gravity is a limit of string theory \cite{Chang:2012kt}.}

\begin{quote}
{\it Any CFT on ${\bf R}\times {\bf S}^{d-1}$ can be interpreted as a theory of quantum gravity 
on an asymptotically $AdS_{d+1} \times {\bf M}$ spacetime, where ${\bf M}$ is a compact manifold.}
\end{quote}
(The reason to put the CFT on ${\bf R}\times {\bf S}^{d-1}$ instead of ${\bf R}^{d-1,1}$ is to replace the Poincar\'e patch of AdS that arises in the near-horizon p-brane limit with its geodesic completion, global AdS, that has ${\bf R} \times {\bf S}^{d-1}$ as its boundary.)
 The utility of this statement is limited by the fact that for a general CFT, the quantum gravity side is not independently defined. 
  
In the eighteen years since its formulation, there have been some 10,000 follow-up papers on AdS/CFT. The canonical review article from the early days is \cite{Aharony:1999ti}; a  recent one is \cite{Hubeny:2014bla}. Here are a few comments.

\begin{enumerate}
\item Gauge/gravity duality is a conjecture awaiting proof in even the most well-understood cases. But there is enough highly nontrivial evidence (e.g. the supersymmetric index  \cite{Kinney:2005ej}, exact results from integrability \cite{Beisert:2010jr} and phenomenological evidence), that it seems not unreasonable to classify it as `true but not proven'. The argument has been made that if such a ridiculous claim were not true, it would have been noticed by now!
\item As mentioned above, part of the duality statement is an identification between dimensionless parameters. In the bulk, there are two important numbers, the ratio of the AdS radius to the Planck length $\ell_{AdS}/\ell_P$ and the ratio of the AdS radius to string length $\ell_{AdS}/\ell_s$. The first ratio governs the ability of the bulk to support coherent excitations of many gravitons without forming black holes. The second governs string-scale locality. Both of these must be large for Einstein gravity to approximately hold in the bulk. The Planck length is automatically less than the string length if the string coupling $g_s$ is small. The condition $\ell_{AdS} \gg \ell_s \gg \ell_P$ is the semiclassical limit. 
	
	Thus, a minimum requirement for a CFT to support a semiclassical bulk dual is that it has two large dimensionless parameters as well. Moreover, one of the parameters ought to be a large coupling constant because free field theory contains many high spin conserved currents which would be dual to light high spin fields in the bulk \cite{Klebanov:2002ja}.  To get something resembling a low energy effective theory of spin-2 gravity, we must generate large anomalous dimensions for these operators.  
	
 In the case of $N=4$ SYM, the two parameters are $N$ and $\lambda$, as in Figure 1.1.
 
 What the exact conditions are for a CFT to admit a semiclassical bulk dual is an open question.
	
\item An extensive dictionary \cite{Gubser:1998bc, Witten:1998qj} was developed to use the gravity theory to study strongly coupled CFT's, in the semiclassical limit. To list a few of the entries,
\begin{enumerate} \item The generators of the conformal group of the CFT match the generators of AdS isometries. In particular, the CFT Hamiltonian that generates translations along the ${\bf R}$ of ${\bf R} \times {\bf S}^{d-1}$ matches the asymptotic generator of time translations in global AdS. This means that thermodynamic quantities (energy, entropy, temperature) of the CFT and the equivalent global gravitational quantities defined at asymptotic infinity agree as well.
\item Low dimension local operators in the CFT map to light bulk fields, with the expectation value of the operator setting the asymptotic boundary conditions for the field. 
 Computation of correlators in the 
 CFT are dual to scattering problems in the bulk. 
\item Expectation value of Wilson loops in the Euclidean CFT \cite{Maldacena:1998im} and entanglement entropy for spatial subregions on a constant-time slice in the CFT \cite{Ryu:2006bv, Ryu:2006ef} are described by areas of homologous minimal surfaces in the bulk. \footnote{Note that in both cases, we can compute the exact value of the CFT observable for certain shapes, at all values of ($N$, $\lambda$) in $N=4$ SYM and some other CFT's \cite{Pestun:2007rz, Kapustin:2009kz, Lewkowycz:2013laa}; perhaps this is a hint to understand the bulk away from the semiclassical limit, though as just some function of ($N$, $\lambda$) it seems hard to interpret.}
\item Thermal states of the CFT, at temperatures above the Hawking-Page transition, are described at leading order by AdS-Schwarzschild black hole solutions to the Einstein equations in the bulk \cite{Witten:1998qj}. One can understand this quantitatively by checking the dominance of the AdS black hole over the thermal gas in the saddle point approximation to the Euclidean path integral. Qualitatively, the intuition is that putting AdS at a finite temperature is like coupling it to an external heat reservoir where energy sloshing around in the bulk has some probability to form a large black hole, then because the AdS geometry is like a box, large enough black holes will recapture their Hawking radiation reflected from the boundary, and are stable.  The classical black hole background was famously used to study properties of the quark-gluon plasma at RHIC.

\end{enumerate}

However, these dictionary entries have in common that the arrow of information goes from bulk to boundary. The questions we ordinarily ask of CFT's turn into bulk computations that are anchored to the asymptotic region. 

\item The reverse problem, how to describe physics in the bulk interior using the CFT, is very poorly understood. A longstanding dream is to use AdS/CFT to understand the quantum properties of black holes, but at a less ambitious level, non-singular gravitational systems like neutron stars which are configurations of bulk fermions that backreact on geometry,
or indeed any localized phenomena in our universe, should also have embeddings in an asymptotically AdS spacetime and hence a description in CFT. However, we cannot engineer such bulk systems because we do not even know how to define a local bulk operator in AdS in a self-contained way in the CFT. 
Studies of the bulk away from empty AdS generally have assumed (solutions to) the classical Einstein or supergravity equations.

 Conversely, any excited state of the CFT is dual to some bulk state with backreaction on the AdS geometry. Given a state of the CFT, we can ask what geometry it describes. The effects of backreaction are classically suppressed by the 10d Planck constant which scales as $1/N^2$ in AdS units. States with energies parametrically smaller than $\mathcal{O}(N^2)$ are well approximated by QFT/classical string theory on the AdS. For states with finite backreaction,  holographic renormalization \cite{Balasubramanian:1999re, Skenderis:2002wp} tells us how to continue CFT expectation values of the stress tensor and other operators in a power series expansion into the bulk using the classical equations of motion. But the generic evolution leads to a naked singularity in the bulk \cite{Hubeny:2006yu}. In some sense, this pathology is not surprising because we know the classical EOM's have string/quantum corrections that accumulate under evolution. Indeed, the converse of the dictionary entry 3(d) on the previous page is that a generic CFT state with energy of order $\mathcal{O}(N^2)$ is a microstate of a bulk black hole, where we certainly expect large corrections to classical gravity.

\end{enumerate}

\subsection{Dualities in field and string theory}

Let's take a step back from AdS/CFT and put it in context among the other QFT and string dualities. \footnote{See \cite{Seiberg:2006wf}, \cite{Polchinski:2014mva} for an enumeration of quantum dualities.} 
To organize our thinking, we can start by asking what it is we hope to gain from finding dualities. Taking another step back, we recall the mission statement from the overview, which is to gain mastery over the Hilbert space of quantum gravity. 

This Hilbert space consists of all excitations atop a vast moduli space of string vacua. The moduli space is often drawn as Figure 1.2. 

\begin{figure}[h]
\centering
\includegraphics[width=0.65\textwidth]{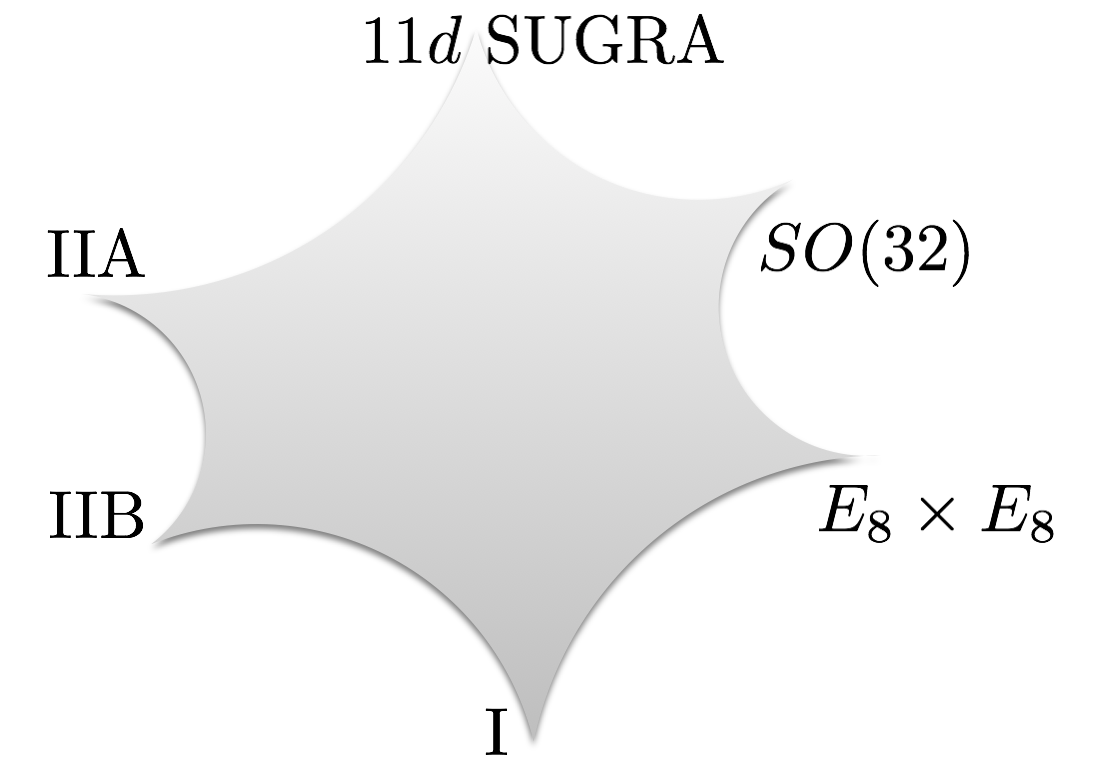} 
    \caption{A slice through the moduli space of string/M-theory. The moduli include the string coupling $g_s$ and the ratio of the string scale to the characteristic length scale of spacetime, $\ell_s/L$. The corners in the drawing represent regions where various descriptions are weakly coupled. As the theory gets strongly coupled in one description, it may become weakly coupled in another one.}
\end{figure}

Because AdS/CFT is a QFT/string duality, it is useful to keep another abstract space in mind, the space of QFT's. Each point on this space is a full-fledged theory with an attached Hilbert space of its own. 
We can also put a natural vector field on this space, that is the RG flow.
At least historically, one might think of QFT space as a specification of the number and types of quantum fields in the UV and their interactions. For example, Figure 1.1 would be a subspace thereof. This picture captures the qualitative idea though the naive description is both redundant and incomplete. \footnote{A ``modern" point of view is that QFT's need not have Lagrangian descriptions. On the other hand, some QFT's specified by different Lagrangians are dual. See \cite{pmath} for a list of grievances about the naive Lagrangian formulation of QFT.}  In terms of our goal which is to understand the string Hilbert space, the QFT space is an auxiliary construct that will help us shed light on it.

The string coupling $g_s$ is a modulus. This leads to the first desirable feature that we might ask for in a duality: strong-weak coupling. We don't understand ordinary quantum theories that are strongly coupled, much less strongly coupled quantum gravity. But we understand quantum theories well enough at weak coupling, which is equivalent to the theory having a classical limit. S-dualities allow for an abstract space to be covered in ``coordinate patches" of weakly coupled descriptions that we have control over. Of course, even with a weakly coupled description, we don't expect to solve the dynamics of the theory around any random state in its Hilbert space, but we can compute perturbatively around the vacuum. There are many examples of S-duality in ordinary, particularly supersymmetric QFT's.
%
%
%
In string/M-theory, S-duality acts on the moduli space where it famously relates the descriptions that label various corners of Figure 1.2.

A second desirable feature that one might want in a string theory is a notion of string-scale locality or a supergravity limit. Even if the theory is weakly coupled so that it makes sense to partition physics into a spacetime background and excitations atop it, there is still the question whether the spacetime is large in string units so that excitations in one region don't affect those in another. Duality in this variable, that inverts the ratio of the string length $\ell_s$ to the characteristic length scale of the spacetime, is called T-duality, which is an S-duality of the worldsheet theory. As energetic strings spread and become large, there is also a sense in which T-duality acts as a UV/IR duality to recover locality at high energies \cite{Hori:2001ax, Giveon:2015cma}. T-duality also famously relates the theories labeling corners of figure 1.2. 

So in the absence of holography, the situation is the following. There is a very large moduli space for string/M-theory, most of which is terra incognita. Presumably our physical reality is described by a state atop one of the points on the moduli space. If we know that we have a strongly coupled description, there is little we can say. Fortunately, string dualities provide different descriptions for points on this moduli space. But even with a weakly coupled  description, our ability to compute dynamical phenomena is fairly limited in  perturbation theory. Mostly we are allowed to probe a pre-existing, time-independent background. Also in some cases we have tools to compute static features of the theory at strong coupling (i.e. partition functions and their cousins) but these only tell us very coarse, global information about the theory.
%
%
 With this level of technological development it is difficult to attack the great non-perturbative problems in quantum gravity.

Where AdS/CFT fits in this picture is that it maps some subspace of the ``QFT theory space" onto AdS vacua of string/M-theory. There are other examples of non-gravitational theories that are equivalent to non-AdS vacua of the string/M-theory moduli space,  e.g. the BFSS matrix model \cite{Banks:1996vh} and little string theory \cite{Aharony:1998ub}. Part of the space of QFT's covers a part of the string moduli space. The attached Hilbert spaces are identified, where the bulk Hilbert space consists of those finite-energy excitations of the vacuum that keep us in the the superselection sector. An infinite energy excitation, or non-normalizable mode, adds a deformation to the action of the dual QFT. The great qualitative advance that this brings to the quest of understanding quantum gravity is that it tells us each of the nonperturbative questions can in principle be understood in terms of the physics of CFT.

\subsection{Open questions for gauge/gravity duality}
 
To recapitulate, a major hope for AdS/CFT is to use it to draw general lessons for quantum gravity in our universe. 
A dictionary is reasonably well developed in the direction of using classical gravity to study the CFT, but the converse problem how to organize  the information in certain CFT's into a theory of quantum gravity  with a semiclassical limit is hardly understood at all.  

The rest of this section is a naive attempt to classify the open questions in AdS/CFT. Roughly we can categorize into three lines of inquiry. First there are attempts to use the AdS/CFT dictionary in the classical gravity limit to study the strongly coupled CFT as a toy model for real-world systems, such as condensed matter ones and QCD. Gauge/gravity pairs can be engineered from the ``bottom up" so that the gauge theory side exhibits qualitative similarities to the real-world systems. Many things are easier to compute in classical gravity than in strongly coupled field theory. This approach is also useful to shed light on top-down examples of holography where the boundary theory is more mysterious than a CFT.

Then there are questions aimed at better understanding the logic of holographic duality itself. 
(The distinction between this category of problems and the previous/next one is blurry; I'm using it it to contain all the issues that aren't directly motivated by physical applications to the boundary or bulk).
 Here we might ask: 
Which CFT's have duals with semiclassical gravity limits? 
Without a priori assuming AdS/CFT, are there reorganizations of CFT data that give rise to what look like coarse-grained features of AdS geometry, and to what extent do such reorganizations make contact with and teach us about the AdS/CFT duality (in particular: what is the connection to the renormalization group, which has now been explored from many angles \cite{deBoer:1999xf, Heemskerk:2010hk, Vidal:2007hda, Swingle:2009bg, Qi:2013caa})? 
Also, we can continue to write the dictionary one entry at a time. 
Given a gauge/gravity pair, what does any feature of the CFT that we can think of -- be it an observable, some more abstract quantity like entanglement entropy, or an equation/theorem -- translate to on the gravity side? This encompasses a large number of well-defined questions that we might study for its own sake, though physical applications often follow.

Finally, there are the questions directly motivated by major challenges in string theory/quantum gravity, where we are interested in leveraging AdS/CFT as a tool to solve them. 
Already at low energies, we can ask,
\begin{itemize}
\item[$\diamond$] {\it What does AdS/CFT teach us about string-scale geometry?}
\end{itemize}

A general challenge for string theory that goes well beyond AdS/CFT is to explain the nature of spacetime. Is Riemannian geometry an approximation to a new string-scale notion of geometry? Note this is a classical question. 
Intuition strongly suggests yes, a simple argument is that Riemannian geometry is simply the theory of the spin-2 graviton and there are a zoo of higher-spin excitations on equal footing at the string scale.
Are there collections of underlying degrees of freedom whose macroscopic arrangements are what we interpret as conventional spacetime?
%
%
Again, various arguments suggest yes -- chief among them the brane ensemble behind AdS/CFT itself, which hints at a noncommutativity to spacetime at the string scale --
but this needs to be made precise. 
This would-be underlying structure is supposed to go beyond placing branes on a pre-existing flat background: it is somehow to be a precursor altogether  from which the interpretation of spacetime, or indeed many dual interpretations, can emerge 
-- but we are lacking even a qualitative idea what such a formulation might look like. 

AdS/CFT offers a new perspective on this problem, as spacetime along with all its string-scale structure emerges at once from the CFT. 
It's useful to keep three pictures in mind, the CFT, the low-energy effective  gravity description of the bulk with its familiar notions of geometry, and the amorphous bulk string-scale picture underlies it; I'll call this the ``non-geometric bulk picture".


To begin to address the problem, a first step is to understand

\begin{itemize}
\item [$\diamond$] {\it How, in the classical limit of AdS/CFT, does geometry emerge from the CFT?}
\end{itemize}

If we want to go beyond the classical geometry eventually, we must first understand all features of the geometry algebraically: the bulk metric itself should be the output of an algorithm on the CFT data. See \cite{Berenstein:2014pma} for some remarks on how to formulate the problem of emergent classical geometry.

One approach is to ask what features of the CFT give rise to features of geometry in the bulk (areas, volumes, and so on), and then whether we can uniquely reconstruct the bulk metric from these data. In recent years it was suggested that the classical geometry of the bulk is related to entanglement and other information-theoretic quantities in the CFT. However the important step of whether such quantities (and if so, which) can really be inverted to yield the metric everywhere where spacetime is weakly curved, is presently unclear. This will be discussed in Section 1.5.

Another idea is to try to define a local bulk operator in AdS by a definition intrinsic to the CFT. Provided it can be established, the moduli space of the CFT operator is the bulk geometry. For early work see \cite{Banks:1998dd}, \cite{Harlow:2014yka} for a recent review, also the recent constructions \cite{Miyaji:2015fia, Verlinde:2015qfa, Nakayama:2015mva} in AdS$_3$/CFT$_2$. 
	
Note that whatever in the CFT gives rise to the bulk metric, it is natural to expect that its dynamics should give rise to metric dynamics, i.e. the Einstein equations or higher-derivative gravity EOM's, around states with a geometric bulk description where the curvature is small. The existence of a satisfactory explanation is a consistency check of a would-be proposal for bulk reconstruction. The extent to which CFT entanglement dynamics reproduces the Einstein equations will also be discussed below in section 1.5 and some quantitative details reviewed in Chapter 2.

To speculate about the future, if we had an algorithm to recover the bulk metric and its dynamics from a state of the CFT, we can imagine making progress towards formulating string-scale geometry by seeing how that answer depends on $\lambda$, or applying the algorithm to CFT's dual to higher-spin gravity as a warm-up. (E.g. see \cite{Ammon:2013hba, deBoer:2014sna} for studies of how 3d higher-spin gravity generalizes the areas of the Ryu-Takayanagi formula with the vev of a certain bulk Wilson line, that cite understanding what replaces classical geometry in such models as motivation,  although Ryu-Takayanagi is not sufficient to reconstruct the bulk metric, as we discuss below). Also perhaps how collective excitations of the CFT make strings/branes in the bulk (select examples are composite operators in the gauge theory \cite{Witten:1998xy, McGreevy:2000cw, Berenstein:2002jq}) is a clue. But these speculations are of little value before we first understand the emergence of classical geometry from the CFT. 

In principle, a different approach that one can imagine towards emergent geometry in string theory, without attempting to reconstruct the entire bulk of AdS, is to study black holes whose interiors are clean examples of emergent space. 
E.g. if we understood how to work directly with black hole microstates at strong coupling, 
 we could see exactly how the interactions of infalling strings with fractionated branes on the strong-coupling side of the correspondence transition \cite{Susskind:1993ws, Horowitz:1996nw} are consistent with a  geometric interpretation of propagation in a black hole background.
Moreover, perhaps the physics responsible for the black hole interior (see \cite{Martinec:2014gka} for a recent picture) is similar to that of the "non-geometric" picture underlying empty spacetime since empty spacetime can be carved into Rindler wedges (and dS space, in any case, has horizons). 
Of course we don't know how to describe black hole microstates at strong coupling, so this just converts one currently intractable problem into another.



But it leads us to a related question,

\begin{itemize}
\item[$\diamond$] {\it What does AdS/CFT teach us about the nature of horizon entropy in string theory?}
\end{itemize}

 The reason one might hope that AdS/CFT has something to say about it is the resemblance of the black hole entropy formula to the Ryu-Takayanagi formula equating position-space entanglement entropies of the CFT to extremal surfaces in the bulk (see section 1.4 below). In some cases the two coincide, so AdS/CFT already suggests something qualitatively new, which is that horizon entropy (in some examples) should be understood as entanglement entropy, but not of geometrically organized degrees of freedom in the bulk, rather of geometrically organized degrees of freedom on the boundary, which are non-locally organized ones in the ``non-geometric picture" of the bulk. 
 %
 %
 %

On the other hand, extremal surfaces in AdS are in general less symmetric than black holes, so at the quantitative level, perhaps the arrow of information should go the other way. E.g. the proof of Ryu-Takayanagi itself \cite{Lewkowycz:2013nqa} is a generalization of black hole methods at the classical level to situations of reduced symmetry. A recent paper \cite{He:2014gva} applied orbifold calculations of black hole entropy in string theory from the early 90's \cite{Dabholkar:1994gg, Lowe:1994ah} to compute bulk entanglement entropy; perhaps some of the work on black holes in string theory since then (\cite{Giveon:2015cma, Martinec:2014gka, Giveon:2014hfa, Ben-Israel:2015mda} and many others) could shed light on aspects of holographic entanglement as well. I will return to this later; the point is the cross-fertilization between these ideas.


%
%

Another puzzle, about which I have little to say is

\begin{itemize}
\item[$\diamond$] {\it What does AdS/CFT teach us about the breakdown of effective QFT suggested by the firewall problem in a theory of quantum gravity?}
\end{itemize}

See \cite{Almheiri:2012rt} for motivation, \cite{Papadodimas:2015jra} around black hole horizons, \cite{Almheiri:2014lwa} around empty AdS.

As we go to  $\mathcal{O}(N^2)$  energies in the CFT where bulk backreaction becomes of order one, the bulk should start to reveal nonperturbative features of quantum gravity in its full glory. As the CFT remains our only nonperturbative definition of quantum gravity to date (modulo the related examples listed previously), the hope is that we can embed into the CFT all of the nonperturbative quantum gravity puzzles and solve them there. For example, one of the immediate consequences of AdS/CFT was to resolve the information problem in favor of unitary evolution for black holes.
	Naively, trying to understand high-energy bulk problems from the CFT seems premature with as poor an understanding of the low-energy physics as we have now; it seems a lot to ask that we can reconstruct features of local physics at the black hole horizon when we don't understand bulk locality in the AdS vacuum;
	but to give a flavor of the questions we'd someday like to answer, they include: what is the microscopic mechanism by which black hole evaporation preserves information? What is the experience of the infalling observer? What do we learn about emergent time?

\section{Entanglement entropy}
																			
Consider a quantum system with multiple degrees of freedom. Its state is described by a single global wavefunction, so there are correlations between subsystems. Suppose the state is an eigenstate of some operator. Then if we know the value of the operator acting on any subspace, we know its value on the complement as well.
The canonical illustration is the singlet state of two quantum spins $|\Psi\rangle = \frac{1}{\sqrt{2}}(|\uparrow\downarrow\rangle - |\downarrow\uparrow\rangle)$. It is an angular momentum eigenstate with eigenvalue zero. If we know that a pair of spins are in this state, we can fly them to opposite ends of the universe, perform a measurement on one, and ``instantaneously" determine the spin of the other. Clearly, this phenomenon is general. This completely ubiquitous feature of quantum systems goes by the name of entanglement. 

 Suppose the Hilbert space of our system factorizes, and we partition the system into subsystems $A$ and $B$ such that the Hilbert space decomposes into $\mathcal{H} = \mathcal{H}_A \otimes \mathcal{H}_B$. The observer restricted to region $A$ has access only to a subset of the local observables and sees the reduced density matrix $\rho_A = {\rm Tr}_B\,\rho$. The entanglement entropy (EE) of region $A$ is defined as the von Neumann entropy of the reduced density matrix: 
 \be\label{defee}
 S_{EE} = -{\rm Tr}\rho_A \log \rho_A\,.
 \ee
 When the full system is in a pure state, the EE quantifies the number of entangled bits between subsystems $A$ and $B$. When it's in a mixed state, the EE picks up part of the classical (thermal) entropy as well.
The EE is not a good observable, as it's nonlinear in the density matrix. A linear operator that quantifies entanglement cannot exist since entangled states are superpositions of pure states. 

Large amounts of entanglement are a general feature of states in QFT's. For example, the vacuum is an eigenstate of the Hamiltonian, so any partition of the vacuum into subregions is entangled in the sense described above. It's perhaps useful to keep in mind that ``entanglement" is a fancy word for correlations which are more traditionally discussed in the context of QFT's, and the two notions can qualitatively be interchanged. 

In QFT's, the Hilbert space is a tensor product of degrees of freedom at each position, modulo extended operators. Disregarding these (see e.g. \cite{Ghosh:2015iwa} for some progress how to understand EE of Wilson loop operators in lattice gauge theory), subsystems $A$ and $B$ are usually taken to be spatial subregions of the underlying spacetime manifold of the QFT on a constant-time slice, though any ``field-space" partitioning of the degrees of freedom into subsets of independent operators should work as well. 
However, EE across a spatial region is always UV-divergent, with the leading divergence going as the area. This ``area law" comes from the entanglement of UV modes across the boundary. (Incidentally, this makes the EE a useful tool to diagnose the scale of non-locality in a non-QFT \cite{Bianchi:2012ev, Hartnoll:2015fca}.) Because any finite-energy state of a QFT looks like the vacuum state in the UV, the divergence structure of the EE can be completely characterized in the vacuum state, where symmetries can be brought to bear \cite{Liu:2012eea}. 

As a side comment, the universal part of the EE of a ball-shaped entangling region -- the first term in the expansion of the EE in the radius of the ball over the UV cutoff length, that does not diverge and has no extensive contribution -- turns out to be a good $c$-function for the weak form of the Zamolodchikov c-theorem in two, three and four dimensions \cite{Casini:2012ei, Jafferis:2011zi}. That is, if two CFT's are UV and IR endpoints of an RG flow, then $(s^V_d)_{UV} > (s^V_d)_{IR}$\,. The explanation behind this observed pattern, and whether one can construct an interpolating $c$-function for non-conformal QFT's out of the entanglement entropy in all dimensions is an open question. 
%

\subsection{Computing the entanglement entropy in QFT}

Position-space EE in QFT's is hard to compute. The main analytic tool, the replica trick, comes from first observing that $$S_{EE} = -{\rm Tr}\rho_A\log\rho_A = -\lim_{n \rightarrow 1}\partial_n {\rm Tr}\rho_A^n\,.$$
On the other hand, ${\rm Tr}\rho_A^n$ for integer $n$ in the vacuum state of the QFT is the partition function of the QFT on $n$ copies of the Euclidean manifold, glued together cyclically along the entangling domain at Euclidean time $\tau = 0$. (See \cite{hartmanhep} for a brief review of the Euclidean path integral). 
For non-integer $n$, the path integral representation is not well defined. The idea of the replica trick is to compute ${\rm Tr}\rho_A^n$ as the partition function for integer $n$, and then analytically continue the answer to non-integer $n$ near $n=1$ to obtain the EE. In practice, this computation is often rather difficult, and it is not always obvious that the analytic continuation is correct. It can be done in simple cases and in some special cases where powerful technical tools (e.g. localization \cite{Nishioka:2013haa}, conformal block expansions in 2d CFT's at large central charge \cite{Hartman:2013mia}) apply.

I will quote one well-known result derived by this method early on \cite{Holzhey:1994we}. For a 1+1 dimensional CFT of central charge $c$ and UV cutoff length $\varepsilon$, the entanglement entropy across a single interval of length $L$ is

\be\label{2dee}
S_{EE} = \frac c 3 \log \frac L \varepsilon\,.
\ee

Another method to compute a special case of position space EE in QFT was pointed out by Casini, Huerta and Myers \cite{Casini:2011kv}. Suppose the entangling subregion is a ball ${\bf S}^{d-1}$ in the vacuum state of a $d$-dimensional CFT. The causal development of the region inside the ball can be conformally mapped to a hyperbolic space ${\bf H}^{d-1} \times {\bf R}$ by a conformal transformation, that takes vacuum correlators of the former to thermal ones in the hyperbolic space. Both the curvature of the hyperbolic space and its inverse temperature are set by the radius of the ball. Therefore the EE of the ball equals the thermal entropy of the hyperbolic space. As this relies only on symmetry, it is an exact result.
	But the answer to the question `what is the thermal entropy of an arbitrary CFT in hyperbolic space' is not known in general.

\section{The Ryu-Takayanagi formula}

However, by far the easiest way to compute entanglement entropies in large $N$ CFT's with holographic duals was discovered by Ryu and Takayanagi (RT) in 2006 \cite{Ryu:2006bv, Ryu:2006ef}. They noticed that the formula \eqref{2dee} for EE's of single intervals in 2d CFT's was identical to the formula for geodesics in  AdS. 

AdS$_3$, in Poincar\'e coordinates, has the metric
\be  
ds^2 = \frac{L_{AdS}^2}{z^2}(dz^2 + dx^2 - dt^2)\,.
\ee
In this geometry, consider two points $x=0, x=L$ on a constant-time slice, on the asymptotic AdS boundary with a cut-off at $z =\epsilon$. The geodesic that joins them has length
\be\label{adsg}
2L_{AdS}\log \frac L \epsilon\,.
\ee
Eqs. \eqref{2dee} and \eqref{adsg} agree if we identify $\frac c 3$ with $2 L_{AdS}$. In AdS$_3$/CFT$_2$ holography, it turns out that $c = \frac{3L_{AdS}}{2 G^N_3}$\, \cite{Brown:1986nw}.
This led Ryu and Takayanagi to conjecture more generally that

\begin{quote} {\it In gauge/gravity duality, in any CFT and state $|\psi\rangle$ of the CFT with an Einstein gravity dual description, the entanglement entropy \eqref{defee} for the reduced density matrix of a subregion $\partial\mathcal{A}$ of the CFT on a constant-time slice is equal to}
\be\label{rt}
S_{EE}(\partial\mathcal{A}) = \frac{A}{4 G_N} + \mathcal{O}(G_N^0)
\ee
{\it where $A$ is the minimal area of the codimension-2 bulk surface that is homologous to $\partial\mathcal{A}$.}
\end{quote}


\begin{figure}[h]
\centering
\includegraphics[width=0.75\textwidth]{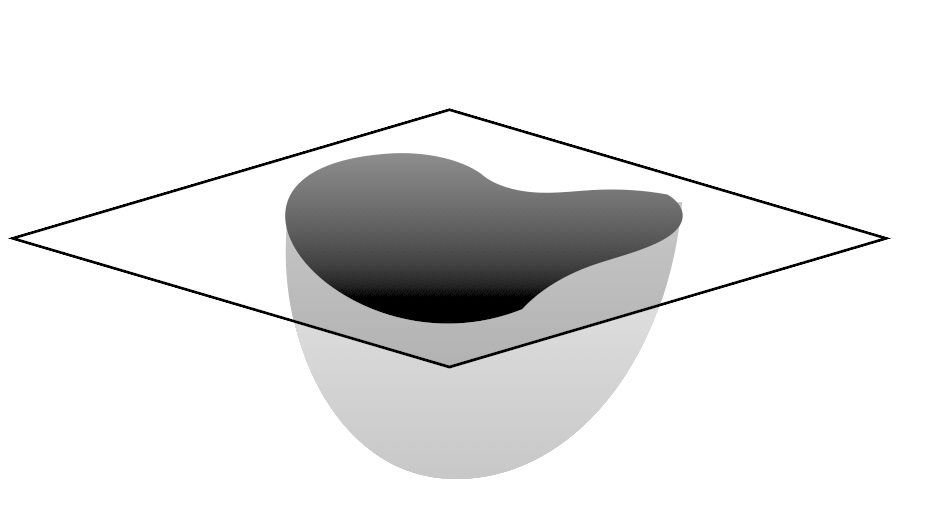} 
    \caption{The Ryu-Takayanagi formula. The entanglement entropy of the  CFT across the shaded boundary region is given by the area of the minimal surface homologous to it in the bulk.}
\end{figure}

For the special case of ball-shaped entangling domains in the vacuum state of the CFT, the Ryu-Takayanagi formula was first proved in  \cite{Casini:2011kv} by exploiting conformal symmetries. Later, Lewkowycz and Maldacena proved it in general \cite{Lewkowycz:2013nqa}. Their work generalizes the following old derivation of the black hole entropy in Euclidean quantum gravity \cite{Banados:1993qp, Susskind:1994sm}. 
In QFT, we know the thermal free energy of the Lorentzian QFT at temperature $\beta^{-1}$ formally equals the partition function of the Euclidean QFT with periodic time $\beta$.
Extending to quantum gravity,
one can
 compute the gravitational entropy of the black hole with Hawking temperature $\beta^{-1}$ by using the thermodynamic identity $S = \beta^2\frac{\partial F}{\partial\beta}$ with $ F = -\frac 1 \beta \log Z$ the free energy, and $Z$ the partition function of Euclidean quantum gravity, subject to the boundary condition that the Euclidean time circle has periodicity $\beta$ at infinity. At leading order in the Newton constant, $\log Z$ is approximated by minus the action of the classical saddle, namely the action of the Euclidean black hole (above the Hawking-Page transition in AdS). The appearance of the partial derivative w.r.t $\beta$ means that we should vary the size of the thermal circle at infinity keeping all other parameters fixed. This introduces a conical deficit in the bulk, so the black hole entropy is the difference between the Euclidean actions of the smooth black hole solution and the configuration with a delta function in curvature at the tip of the Euclidean cigar. One can show that the difference is proportional to the area of the $(d-2)$-sphere at the tip of the cigar, with proportionality constant $1/4G_N$. In this setup, one also can compute $G_N$ corrections as loop corrections to the saddle-point approximation.

Now to sketch the proof (following \cite{Dong:2013qoa}), first recall the $n$-fold covers of the boundary CFT for integer $n$, that we discussed above. We will assume that these are holographically dual to smooth bulk geometries $B_n$ that preserve a ${\bf Z}_n$ symmetry in the bulk. This is the key assumption. The possibility of replica symmetry breaking in the bulk was discussed in \cite{Faulkner:2013yia}. Define the orbifolded bulk geometries $\hat B_n = B_n/{\bf Z}_n$ for integer $n$. They are smooth except at the fixed points of the ${\bf Z}_n$ action, which form a codimension-2 surface. Unlike the replicated boundary, the bulk orbifolds can be sensibly continued to non-integer $n$, although the resulting geometries do not have interpretations as orbifolds of smooth geometries.  The variation of the action of $\hat B_n$ around $n=1$, $\partial_n S[\hat B_n]|_{n \rightarrow 1}$, is the analog of the variation in the Euclidean action around the configuration with Euclidean time periodicity $\beta$ that I described above for the black hole, and it gives the entanglement entropy. 
Finally, by demanding consistency of the Einstein equations, one can show that $\hat B_n$ for non-integer $n$ have also a codimension-2 conical defect that approaches the Ryu-Takayanagi surface as $n$ approaches 1, and that the entire non-zero contribution to $\partial_n S[\hat B_n]|_{n \rightarrow 1}$ comes from the location of the conical defect and is its area, $A/4G_N$.

The generalization of the Ryu-Takayanagi formula to both classical higher-curvature theories of gravity and to bulk quantum corrections is an open problem as of this writing. In the direction of higher-curvature corrections, the answer is known in some cases to be the Wald entropy plus extrinsic curvature terms, that would vanish on the horizon of a static black hole \cite{Dong:2013qoa}. Curiously, the corrections are the same as appear in a generalized second law of black hole thermodynamics for higher-curvature gravities \cite{Wall:2015raa}.
	In the direction of quantum (1/$N$) corrections, the leading correction was argued 
	to be the bulk quantum entanglement across the RT surface, thinking of the bulk as an effective field theory living on a fixed background, with the RT surface as a fixed surface in that spacetime \cite{Faulkner:2013ana}. 
 As mentioned above, it is a natural question whether the RT formula can be better understood directly in string theory. 
	
	A new point of view on the RT formula was recently suggested by Headrick \cite{headrick}. The max-flow min-cut theorem in graph theory says: given a manifold with boundary, let $\{{v}\}$ be set of all vector fields on it that have zero divergence and norm $|v| \leq 1$ everywhere. Then the area of a minimal surface homologous to a codimension 1 region $A$ on the boundary is equal to $\int_A *V$ of the flow $V \in {\bf v}$ that maximizes the integral. The application to the RT formula is obvious. From this point of view, we are asked to optimize over all ways to pack the bulk with threads of a fixed size (implementing the bound on the norm) that a current flows through (implementing the divergence-free condition). If there is more to this than mathematical analogy, it would be interesting to give the current  a physical interpretation.
		
\section{Entanglement and spacetime}

In recent years, various authors have suggested that ``entanglement builds spacetime." 

Depending on the reader's perspective, this may either seem like an extraordinary claim or seem reasonably intuitive. After all, ``spacetime" should be thought of algebraically as a set of consistency conditions on of the degrees of freedom in the theory. E.g. in the bulk effective QFT limit, it manifests as imposing a particular structure on the correlation functions of local operators whether on a constant-time slice, or as causality at different times.
	On the other hand, entanglement is a fancy word for quantum correlation.  
	What is needed is to quantify this -- when are the correlation patterns from entanglement consistent with a classical geometry interpretation (and of what geometry)? The goal of this section is to review the status of these issues.
	
	There seem to be two ideas that were pursued in the literature: (1) that {position space entanglement in the CFT is the organizing principle behind the bulk geometry}, and (2) that {some form of non-geometrically-organized bulk entanglement builds spacetime, which holds beyond holography}. When restricted to holography, the second is a weaker claim. It posits that some form of entanglement of the CFT data,  that need not be spatially organized in the CFT, organizes the bulk geometry.
	I will argue that (1) is easy to rule out so in particular, attempts to derive the Einstein equations from inverting Ryu-Takayanagi do not seem fruitful (in section 1.5.2), review the evidence for (2), and discuss prospects to better understand (2) (in section 1.5.6, 1.5.7). This section is structured as an approximately chronological literature review with comments at the end.

\subsection{Early qualitative arguments}  
The idea `entanglement builds spacetime' originates in two papers by van Raamsdonk in 2009 \cite{VanRaamsdonk:2009ar, VanRaamsdonk:2010pw}. Among his arguments was the following ``disentangling experiment". Take a constant-time slice of a CFT on ${\bf S}^d \times {\bf R}$ that has a holographic dual. Suppose we divide the sphere in half. We label the reduced density matrix of the CFT in each hemisphere as $\rho_L$ and $\rho_R$. In the vacuum, the two halves are highly entangled. By the Ryu-Takayanagi formula, the entanglement can be quantified at leading order in $G_N$ by the area of the bulk surface that bisects empty AdS. Now let us consider the family of states $|\psi(\lambda)\rangle = (1-\lambda)|0\rangle + \lambda (\rho_L \otimes \rho_R)$ in the CFT, that interpolates between the CFT vacuum and the state where the degrees of freedom of the two hemispheres have been completely disentangled. This is a formal definition as it involves disentangling arbitrarily high-energy degrees of freedom, so any of the interpolating states lie outside the Hilbert space of the CFT, but let's proceed. As long as the Ryu-Takayanagi formula holds, the entanglement between two hemispheres continues to be geometrized by the area of a surface bisecting the bulk. This area decreases continuously as one increases $\lambda$. 

At the same time, we can use the mutual information $$I(A,B) = S_{EE}(A) + S_{EE}(B) - S_{EE}(A \cup B)$$
as an upper bound for correlations of operators $\mathcal{O}_A, \mathcal{O}_B$ localized to the respective regions \cite{Wolf:2007aa}. Such correlators fall off as a function of distance. Taking the regions $A, B$ to be the left and right halves and again using the RT formula, the proper distances between the two bulk regions increases with $\lambda$. We see that at least at a qualitative level in this toy family of states, the position-space CFT entanglement is a prerequisite for geometric connectivity. 

\subsection{``Position space entanglement in the CFT builds bulk geometry"}
One suggestion in the literature is that {position space entanglement in the CFT} is more than a prerequisite, it is the organizing principle responsible for the emergence of the dual spacetime in AdS/CFT, in the sense discussed in section 1.2.2. A corollary would be that position-space entanglement dynamics in the CFT should suffice to explain the bulk non-linear Einstein equations in any asymptotically AdS spacetime which is dual to a state of the CFT for which the bulk is geometric. By this I mean that the bulk Einstein equations about any such state will be holographically dual to CFT equation(s) of the form $f(S_{EE}) = 0$, where $f(S_{EE})$ is a function of quantities in the CFT, e.g. expectation values of operators and so on, that includes the geometric entanglement entropy of subregions of the CFT in that state. See section 2.2 below for an example how one such equation explains the linearized Einstein equations around the AdS vacuum. 

Constructions related to bulk reconstruction from position-space CFT entanglement  
include  
``differential entropy" of families of boundary regions \cite{Balasubramanian:2013lsa, Myers:2014jia, Headrick:2014eia} which gives the areas of bulk, not-necessarily-extremal surfaces, 
that need not end on the boundary. It is defined as $S_{\rm diff}(\{I\}) = \sum_{k=1}^n [S_{EE}(I_k) - S_{EE}(I_k \cap I{k+1})]$ where $I_k$ are boundary intervals (in $CFT_2$; boundary strips in higher dimensions) on a constant-time slice, and reproduces the area of the bulk surface that the Ryu-Takayanagi surfaces of the intervals are tangent to, in a continuum limit as $n \rightarrow \infty$. 
%
%
Related work is the exploration of ``kinematic space" \cite{Czech:2015qta}, the space of boundary anchored geodesics on a 2d manifold, with an eye towards reconstruction of the manifold in situations where the tangent space is completely covered by the boundary-anchored geodesics.  \footnote{With an eye towards following paragraphs, note that these constructions can be generalized inside the entanglement shadow using non-minimal geodesics in place of position-space entanglement \cite{Balasubramanian:2013lsa, czechtalk}.}
 See \cite{Porrati:2003na} for work on using geodesics for bulk reconstruction, including 3d counterexamples, that predates Ryu-Takayanagi.

There is actually a simple observation to rule out the idea that position-space entanglement explains the bulk geometry. This is the ``entanglement shadow" \cite{Balasubramanian:2014sra, Freivogel:2014lja}.
 In generic asymptotically AdS spaces that are not empty AdS, even ones without horizons, there are bulk regions that no minimal surface homologous to boundary regions has access to. In a black hole geometry, the shadow is on the order of one AdS radius away from the horizon. The bulk geometry can be made weakly curved there, and we expect the Einstein equations to hold to arbitrarily good precision, yet no RT surfaces reach it.  So the Ryu-Takayanagi formula cannot in general be inverted to yield the bulk metric, and attempts to reconstruct the full Einstein equations from position-space entanglement dynamics fall flat at this level.
 
 Ref. \cite{Balasubramanian:2014sra} defines a generalization of conventional entanglement in the CFT, ``entwinement", whose bulk description has no shadow. Inspired by the observation that internal degrees of freedom are key to emergent spacetime in known examples (i.e. gauged DOF's in matrix models and fractionated DOF's of the long string in models of the black hole interior), the entwinement computes conventional entanglements in a larger, auxiliary theory where the boundary degrees of freedom have been ``un-gauged". In practice, the authors of \cite{Balasubramanian:2014sra} study this in the most naive example possible: they take $AdS_3/\mathbb{Z}_n$ spacetimes where the minimal geodesics of empty $AdS_3$ become both minimal and non-minimal winding ones, and un-gauge the boundary theory taking $n$ copies of the dual CFT Hilbert space. \footnote{See \cite{Mollabashi:2014qfa, Karch:2014pma} for other proposals of holographic entanglement between internal degrees of freedom in the boundary CFT.}
  Also there is no shadow for holographic R\'enyi entropies. \footnote{Thanks to Xi Dong for pointing this out.}
 These probes are examples of the next general framework that we will explore, namely that entanglement in or information-theoretic organization of the CFT data {\it not} precisely organized by spatial subregion in the CFT is the principle behind the emergence of the dual spacetime. First we discuss the example of the two-sided AdS black hole. 
  
\subsection{The two-sided black hole} 
Consider two copies of a CFT with a holographic dual, each living on its own spatial ${\bf S}^d$. 
Suppose we put the CFT's into the thermofield double (TFD) state 
\be\label{TFD}
|\Psi(\beta)\rangle = \frac{1}{\sqrt{Z(\beta)}}\sum_n e^{-\beta E_n/2}|n\rangle_1 |n\rangle_2\,.
\ee
The sum runs over all states in the CFT's, and $Z(\beta)$ is the partition function at temperature $\beta^{-1}$ of one copy of the CFT. \eqref{TFD} is a pure state in the doubled system. However, if we trace out either one of the CFT's, the reduced density matrix in the other is $\rho = \sum_n e^{-\beta E_n}|n\rangle\langle n|$, the thermal density matrix in that CFT. The two copies of the CFT are entangled in \eqref{TFD} and the entanglement entropy is the thermal entropy on each side.  

Maldacena proposed that, in the semiclassical approximation, the TFD state \eqref{TFD} of two copies of a CFT is holographically dual to gravity on the extended AdS-Schwarzschild black hole background \cite{Maldacena:2001kr}. 
%
%
The justification was
an extension of the AdS/CFT hypothesis to describe Schr\"odinger picture states set up by Euclidean path integrals. The usual holographic hypothesis is $Z_{grav}[\partial\mathcal{M} = \Sigma] = Z_{CFT}[\Sigma]$, or in words, the partition function of quantum gravity in spacetimes $\mathcal{M}$ that asymptote to boundary $\Sigma$ is the partition function of the dual CFT on $\Sigma$. Instead of applying it to the full spacetime, it is applied here to a Euclidean geometry $\tilde \Sigma$ with an open cut at $\tilde\sigma$, that sets up some state in the CFT on spatial $\tilde\sigma$. We conjecture that the boundary state set up by this procedure is dual to the Hartle-Hawking wavefunction that is set up by the Euclidean path integral on the dominant saddle whose boundary is $\tilde \Sigma$, with open cut $\tilde\sigma$. Then we evolve both sides in Lorentzian time. 

%
%

Specifying to the  two-boundary case (though see \cite{Skenderis:2009ju, Balasubramanian:2014hda} for studies of examples with multiple asymptotic boundaries), the TFD state \eqref{TFD} of two CFT's on ${\bf S}^d$ is set up by the Euclidean path integral on $\tilde\Sigma = {\bf I}_{\beta/2} \times {\bf S}^d$ with open cuts at the ends of the cylinder. On the other hand, the Euclidean continuation of the AdS-Schwarzschild black hole has boundary ${\bf S}^1_\beta \times {\bf S}^d$. Its $\tau = 0$ cross section has the boundary ${\bf I}_{\beta/2} \times {\bf S}^d$ with cuts at the two ${\bf S}^d$'s.

The evolution in Lorentzian time should be clarified. There are two times on the boundary, one for each CFT. They can be evolved independently and the action on the boundary state is completely well defined. In particular, the TFD state is an eigenstate of the difference of the two Hamiltonians, $H_L - H_R$. The associated time coordinate is the global Killing time $\tau$ that is spacelike on the initial slice. Another commonly discussed time $t$ that we specialize to below is when both boundary times run from past to future associated to the asymptotic Hamiltonian $H_L + H_R$. 
Note that 
there is no way to uniquely assign a spatial slice of the bulk to each boundary time. Instead we associate the entire spacelike separated region from the boundary points at that time. 

\begin{figure}[h]
\centering
\includegraphics[width=0.4\textwidth]{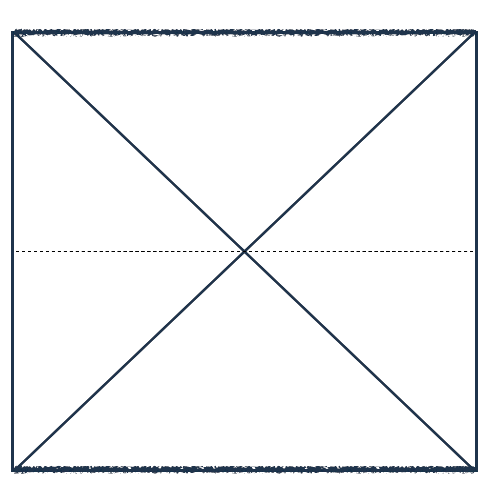} 
    \caption{Penrose diagram for the eternal black hole. The diagram shows the radial and time directions. At each point on the diagram, there is a sphere. The sphere shrinks towards the top and bottom edges of the diagram which are spacelike singularities. The dotted line is the $t=0$ section which has the geometry of an ER bridge.}
\end{figure}

The Penrose diagram of the eternal black hole is shown in Figure 1.4. It has two asymptotically AdS regions that look like the exterior regions to a AdS-Schwarzschild black hole. Observers in the two regions are out of causal contact as long as they stay outside the black hole. However, they can meet inside the black hole. Also, the correlations between the left and right regions are nonzero. 
The $t=0$ slice of the geometry is an Einstein-Rosen (ER) bridge that connects the exterior regions. If we stayed on the $t=0$ slice of the geometry and went from one asymptotic region towards the horizon we would find ourselves in the other region the moment we went through the horizon. But taking time to evolve upwards on both boundaries, the wormhole is not traversable; it grows too rapidly with time.

(As an aside, it was suggested that this fact is a challenge to the idea that (any form of) boundary entanglement is responsible for the bulk geometry. 
For this choice of boundary time, the length of the ER bridge grows indefinitely. One can ask what this geometrizes in the dual QFT, since the vertical entanglement saturates after a scrambling time \cite{Hartman:2013qma}. 
	 We need some other quantity in the gauge theory to serve as a clock after that. Susskind \cite{Susskind:2014rva, Susskind:2014moa} argued that quantity is the the computational complexity of the state in the QFT. Stanford and Susskind conjectured in \cite{Stanford:2014jda} that complexity is dual to the wormhole volume. The conjecture was refined in \cite{Brown:2015bva} to ``complexity is dual to the action of the Wheeler-DeWitt patch". However, these conjectures are confusing because 
	 the would-be dual patch always contains finite support in a strong-curvature region of the bulk.)

The TFD/two-sided black hole duality has interesting implications. Firstly, it's a throwback to an old idea that black hole entropy is entanglement entropy \cite{Bombelli:1986rw, Srednicki:1993im}, although that suggestion seemed to run into problems early on (e.g. \cite{Kabat:1995eq}, references in \cite{Cooperman:2013iqr}, \cite{Solodukhin:2011gn}). Here this indeed appears to be true, but the black hole entropy is not the entanglement of bulk degrees of freedom separated by the horizon or otherwise geometrically organized in the bulk. Rather, it is an entanglement entropy in the dual CFT, that corresponds to some non-local, non-geometrically-organized entanglement of the bulk degrees of freedom. 

Next: the vacuum $|0\rangle \otimes |0\rangle$ in the doubled Hilbert space is dual to two disconnected empty AdS's.  
	Let us consider the TFD state \eqref{TFD} from the point of view of the disconnected AdS's. It's a sum over products of pure eigenstates of each boundary Hamiltonian, which can be described by individual asymptotically AdS spacetimes with a corresponding amount of energy in the bulk  (i.e. partitioned somehow along the fundamental degrees of freedom in the ``non-geometric picture"). At high energies these are the microstates of bulk black holes. So we can interpret the TFD state as either the superposition of product black hole microstates, i.e. maximally entangling two black holes, or as the empty, connected wormhole geometry. We are replacing a more complicated algebraic description with a geometric interpretation. It is just a reorganization of the state into what we decided to call the background and what we call excitations on top of it. 

Granted this fact is true in disconnected AdS's, and assuming the degrees of freedom that constitute black hole microstates are localized in the bulk it is a small step to apply it to entangled black holes in two far-away regions of the same spacetime. To summarize:

\begin{quote}
(``Weak form of ER = EPR"): {\it Two maximally entangled black holes are connected by an ER bridge.}	
\end{quote}

To me it seems the two-sided AdS black hole and its seemingly reasonable extension to entangled black holes more generally, is the only real piece of evidence for ``entanglement builds spacetime" which is on a semi-quantitative footing (along with the circumstantial evidence that `entwinement' is the only bulk probe to date with no shadow).
All the evidence above and below is either not quantitative or isomorphic to (perturbations around) it.

\subsection{AdS-Rindler embedding}

By running the above construction for two CFT's on hyperbolic space instead of on spheres, we can pull the two-sided black hole construction into a single copy of global AdS. Consider the Euclidean path integral on $\tilde \Sigma = {\bf I}_{\beta/2} \times {\bf H}^d$ with open cuts at the two ${\bf H}^d$'s. This gives the thermofield doubled state \eqref{TFD} of the two CFT's on hyperbolic space. By the above hypothesis, it is holographically dual to the two-sided hyperbolic AdS black hole whose Euclidean geometry has the boundary ${\bf S^1_\beta} \times {\bf H}^d$. 

On the other hand, a massless two-sided hyperbolic AdS black hole can be conformally mapped to empty global AdS, \footnote{As a corollary, the two-sided black hole has no ``shadow" if we consider two-sided boundary entangling domains that are supported on both ${\bf S}^d$'s.} where each exterior region maps to a Rindler wedge of the AdS \cite{Emparan:1999gf} and the black hole interior fills the rest. 
Actually, we already encountered the boundary conformal transformation that implements this. It is the inverse of the CHM map \cite{Casini:2011kv} that takes a thermal state of a single CFT on hyperbolic space to the reduced density matrix of its vacuum state on a sphere. 
%

There are arguments for a ``subregion duality" \cite{Bousso:2012sj, Bousso:2012mh, Czech:2012bh} between a Rindler wedge of the boundary CFT and a Rindler wedge in the bulk AdS. One way to state the question of subregion duality \cite{Czech:2012bh} is, given any state of the CFT with a geometrical bulk description, and given a subregion of the CFT, what part of the bulk geometry can be reconstructed from knowledge of the reduced density matrix? Ref. \cite{Czech:2012be} argued that from the density matrix of a Rindler wedge of the vacuum CFT, one should be able to reconstruct the geometry of an AdS-Rindler wedge in the bulk. A slightly different phrasing 
\cite{Bousso:2012mh} is, given any state/subregion with the above criteria, in the classical limit around such a state, what part of the bulk geometry has the property that the local operators in it are dual to CFT operators supported on that subregion? The HKLL, or AdS-Rindler reconstruction \cite{Hamilton:2006az} argued that a bulk operator contained in the {causal wedge} of a boundary region $A$, which is the bulk region enclosed by the minimal geodesic (Ryu-Takayanagi surface) on a constant time slice, can be reconstructed by CFT operators on $A$. \footnote{For multipartite regions, perhaps the appropriate region is the ``entanglement wedge" \cite{Headrick:2014cta, Almheiri:2014lwa, Pastawski:2015qua}} There are subtleties \cite{Bousso:2012mh, Almheiri:2014lwa} (e.g. the same bulk operator would have to be described by many inequivalent boundary operators with support on different regions of the CFT).

%

%

 Since a Rindler wedge of the vacuum AdS is the exterior of a two-sided hyperbolic black hole, an exact subregion duality would be like arguing that in the eternal black hole, either one of the CFT's in isolation describes just the geometry outside the horizon, 
	and the interior is completely a consequence of the entanglement pattern. 
	It is equivalent to a very strong form of ``position-space entanglement builds spacetime" around the AdS vacuum: that in one copy of the empty AdS, entanglement is completely responsible for the existence of any spacetime at all behind a pair of bulk Rindler horizons, which any random point in AdS can serve as a bifurcation point for! Precisely understanding the subregion duality, particularly away from the vacuum,
		seems interesting to clarify the validity of the position-space entanglement picture, that as discussed in 1.5.2, seems approximately true around the AdS vacuum, but not in general. 

\subsection{Beyond AdS/CFT: Gravitational inequalities and relative entropy }

This subsection is a bit orthogonal to the rest of this chapter. I leave it here for completeness, but the reader following the logical flow of previous sections can skip to 1.5.6.

There are proofs of longstanding gravitational entropy bounds using relative entropy inequalities. These don't rely on holography. Roughly, they bound entropy (number of states) in a region by energies or areas, constraining the growth of states
 in systems coupled to gravity, where we can make black holes. See \cite{Bousso:2002ju} for a nice review.

One example is Casini's \cite{Casini:2008cr} reduction of the Bekenstein bound \cite{Bekenstein:1980jp} to the positivity of relative entropy. The Bekenstein bound is the proposal all systems should satisfy \be\label{bek} S \leq 2\pi RE \ee
where $S$ and $E$ are the entropy and energy of the system confined to a region of size $R$. It came from a thought experiment where a probe is lowered into a black hole. Suppose the probe of entropy $S$ and energy $E$ released from a distance $R$ above the horizon, falls into the black hole. We compute the variation of the black hole entropy by converting the energy swallowed by the black hole to the change in its mass to the change in its entropy, then argue $\Delta S_{BH} \geq S$. Eq. \eqref{bek} has no Newton constant in it, so we should be able to interpret it in flat space. The problem was that it was not clear how to define the various quantities. An easy counter-example to the naive interpretation that $S$ is the microcanonical count of states for a QFT in a cavity of size $R$ is if we take the number of species of quantum fields to be large. But if the system is a region in a larger space, we can't localize fluctuations to the region anyway. Casini showed that a natural interpretation of the bound is as a special case of the positivity of relative entropy, which is an inequality that holds for all quantum systems (that will be described in the next chapter). The entropy in \eqref{bek} is interpreted as the change in the entanglement entropy of degrees of freedom in the region with the rest of the space, between the excited state under consideration and  the vacuum state.

Another result is the recent proof \cite{Bousso:2014sda, Bousso:2014uxa} of the Bousso bound \cite{Bousso:1999xy, Flanagan:1999jp}
\be\label{bousso}
\Delta S(B) \leq \frac{\Delta A(B)}{4G_N}
\ee
in the weak gravity limit where the change in the area is first order in $G_N$.  $S[B]$ is the entropy on a light-sheet of the surface $B$ and $A(B)$ is the area of $B$. A light-sheet of $B$ is a null hypersurface that is orthogonal to $B$ and shrinks going into the future. There was also previously a question how to define the notion of entropy $S(B)$ on a light-sheet (for essentially the same reasons). 
%
The recent proof again proposed to sharply define $S[B]$ as the difference in the von Neumann entropy between the vacuum and excited states, as seen by the algebra of operators on the light-sheet, and to use the positivity of relative entropy as the main technical tool. Here the relative entropy related the von Neumann entropy to an energy flux of the modular Hamiltonian, that focuses light rays by the Raychaudhuri equation to constrain the area of light-sheets. 

A related recent paper by Jacobson \cite{Jacobson:2015hqa} 
suggested that the nonlinear Einstein equations in any (not necessarily asymptotically AdS) spacetime follow from a stationarity condition on the entanglement entropy of infinitesimal causal diamonds in spacetime.
Despite superficial similarities, the argument is quite different from the AdS/CFT derivation of the linearized Einstein equations from entanglement thermodynamics reviewed in Chapter 2 below,
since it applies the entanglement first law not in the dual CFT but directly in the gravity side. Accordingly, it requires some assumptions that short-distance behavior in a full-fledged quantum gravity theory resembles that of a CFT vacuum.


\subsection{ER = EPR}

``ER = EPR" \cite{Maldacena:2013xja} is the conjecture that any pair of entangled qubits is dual to a non-geometric ER bridge. One might say

\begin{quote}
(``Strong form of ER = EPR"): {\it Maximally bipartite-entangled bulk degrees of freedom are equivalent to  a non-geometric version of the ER bridge; bulk entanglement builds spacetime.}
\end{quote}

It is the same idea as the weak form stated in 1.5.3, but we replace ``black holes" by more general objects and the geometric ER bridge by some non-geometric one. This definition is not very precise, the idea will hopefully become clearer over the course of this section.

Some preliminary comments: 
\begin{enumerate}
\item The claim of \cite{Maldacena:2013xja} is {\it not} that any entanglement, even strong bipartite entanglement (such as the TFD state \eqref{TFD} in a dual boundary description of two CFT's acted on by some unitary matrix, so that the entanglement entropy between the CFT's is the same as in the TFD state itself) should be identified with a {\it geometric} ER bridge, which was challenged e.g. in \cite{Marolf:2013dba, Balasubramanian:2014gla}.

\item The proposal does not imply nonlinearity of quantum mechanics; although entanglement is not an observable in quantum mechanics, neither is the (non)existence of a wormhole in quantum gravity -- this was emphasized early on by Motl \cite{lumo} and again recently in the paper \cite{Bao:2015nca}.

\item When pulled onto the boundary in AdS/CFT, a statement about entangled bulk DOF's becomes one about the entanglement of the boundary DOF's dual to the bulk ones, where we expect the map between bulk and boundary to be nonlocal;  hence a naive reading of ER = EPR suggests that bulk geometry arises from boundary entanglement, but that need not be organized by position space on the boundary. 
 \end{enumerate}

There is no independent definition for the ``non-geometric" ER bridge so it's hard to make sense of this proposal as stated. 
The goal here is to see how far we can get.
Let's start by reviewing some motivation and arguments in the literature for it. There should be a formulation in string theory,  I discuss this briefly later in this section.

\begin{enumerate}
\item The original motivation was to address the firewall problem.  In brief, the firewall \cite{Almheiri:2012rt} is a refinement of the black hole information problem that exposes an apparent inconsistency between unitarity, 	
	the equivalence principle, 
	and low-energy effective field theory.  
 It goes something like this. Given that AdS/CFT exists, let us suppose black hole evaporation is unitary. Consider the von Neumann entropy of the Hawking radiation. At early times it is zero (there is no radiation) and at late times it is zero (by unitarity), so the von Neumann entropy over time must be an `inverted V' shape; after some time called the Page time, the radiation that comes out of the black hole must be entangled with earlier radiation, to purify the final state. On the other hand, assuming effective field theory, quantum fields in the vacuum are maximally entangled across the black hole horizon. By monogamy of entanglement, these things can't simultaneously be true. The ``firewall" is defined as the  absence of the usual QFT entanglement pattern across the horizon. 

Naively, one way out that is logically consistent with monogamy of entanglement and no firewall ($`` A = R_B"$) is simply to declare that the interior mode of the black hole after the Page time is the same as an early radiation quantum. Then both can be maximally entangled with late radiation as they are secretly the same. But this leads to nonsense. 
Maldacena and Susskind proposed a refinement of this idea,``$R_B \rightarrow A$" \cite{Susskind:2013lpa, Susskind:2014moa}, arguing that Hawking radiation is connected to the black hole interior by non-traversable wormholes. Macroscopically, suppose we collect all the Hawking radiation at the Page time and collapse it into a second black hole. Then the pair of black holes have the causal structure in Figure 1.4, with some identifications between the left and right sides since there is only one asymptotic region. This is the ``weak form" of ER = EPR discussed in section 1.5.3, that follows from the two-sided AdS black hole holography. The strong form just argues there is some continuity before and after we collapsed the Hawking radiation. While not identifying $A$ and $R_B$, it allows for causal dependence between them. This resolves the three-way monogamy of entanglement conflict, which assumed the modes were independent. 


	\item In AdS/CFT, a self-consistent picture appears when one studies entangled heavy quarks in the boundary CFT \cite{Jensen:2013ora}. The quarks are dual to a bulk string connecting them. The authors of \cite{Jensen:2013ora} showed that classically, the induced metric on the bulk string worldsheet contains the same causal structure as that of the eternal black hole, and argued that the area of the induced worldsheet horizon is the entanglement entropy of the quarks. The Schwinger instanton responsible for quark pair creation is similar to the Euclidean instanton that generates the Hartle-Hawking state in the eternal black hole construction \cite{Sonner:2013mba}. On the other hand, the worldsheet is inheriting this structure from the Rindler decomposition of the bulk. 
			
	\item Susskind has some nice cartoon arguments, though they're logically equivalent to what is written above, just presented a bit differently. One is a cartoon to illustrate the continuity \cite{Susskind:2014moa}. Aside from pair creation that features in the above example, another process that generates two entangled systems is a ``fission" where a black hole decays into two smaller ones, with exponentially small probability. If it decays into two equal-sized ones, we are back to the two-sided black hole of Figure 1.4 (in the same space). But the black hole can also split  asymmetrically, indeed with higher probability, into a large component and small component.  The limit as the small component is replaced by an elementary particle is the claim of ER=EPR. \\ A separate argument \cite{Susskind:2014yaa} asks us to contemplate the difference between the following three scenarios: (i) Global AdS with two clouds of maximally entangled particles, (ii) global AdS with those clouds of particles collapsed into large entangled black holes and (iii) global AdS with two large unentangled black holes. As long as Ryu-Takayanagi holds in cases (ii) and (iii), the difference between the black holes being entangled and not must be reflected by the area of the minimal surface. But the surface bisecting the bulk would appear unchanged unless the entanglement added a ``handle" to the global topology of AdS, which is the ER bridge. This is pretty much identical to the argument in section 1.5.2. 
		But now consider scenarios (i) vs (ii). In scenario (i), the individual backreaction of each particle on the bulk is small, and entanglement of the clouds is captured by the subleading FLM correction to the Ryu-Takayanagi formula \cite{Faulkner:2013ana}, as bulk entanglement across the bisecting surface. But we can freely go between (i) and (ii) by collapsing the clouds into black holes or letting the black holes evaporate. 
		\item
		
	As a bit of a side comment, to check when entanglement gives rise to {\it geometric} wormholes (though as mentioned above, this is not quite the main point of the conjecture), a check one can do is to take AdS/CFT with two boundary CFT's in entangled states other than the TFD one \eqref{TFD} and see when the two-sided correlations are large. \footnote{However, see \cite{Shenker:2013pqa, Shenker:2013yza, Leichenauer:2014nxa} for examples where there is a ``long semiclassical wormhole" in the bulk for which two-sided correlations are small  (though the motivations for those studies is different). 
} 

Papadodimas and Raju \cite{Papadodimas:2015jra} performed a similar computation as an application of the main thrust of their work, which was to construct (state-dependent) local bulk operators behind the black hole horizon. 
	Suppose we had any entangled state in the doubled Hilbert space of two CFT's that was a high energy eigenstate of the right Hamiltonian $H_R$, so from the asymptotic right boundary, it looks like a black hole. Moreover, suppose that in the classical limit we knew how to define local bulk operators behind the black hole horizon in terms of CFT ones. Then we could check if the two-point function of an operator behind the right black hole horizon with operators in the left CFT is nonzero, suggesting structure reminiscent of the two-sided geometry (Figure 1.4). 
		They computed such correlators for various two-sided states 
			and noted when the correlators in each case were order one or exponentially small. They did not attempt to reconstruct bulk geometry from the correlators. 			

	\end{enumerate}
	
	To summarize, the logic behind the strong form of ER = EPR is to take for granted the weak form that large maximally-entangled black holes are joined by an ER bridge, in a weak curvature regime where both structures make sense geometrically, 
	 then argue by continuity. 
	The discussion clearly suffers from there being no definition for a ``non-geometric wormhole." 
	 We would like to make the statement more precise.
	  In the rest of this section, I will speculate in this direction.
	  
	To take stock of where we are, there are two (related) questions of obvious importance:
	
	\begin{itemize}
	\item[$\diamond$] What is the definition of the ``non-geometric ER bridge?" What are its properties?
	\item[$\diamond$] If ER = EPR explains the emergence of spacetime, what is the mechanism by which sufficient numbers of the ``non-geometric wormholes" (also perhaps analogous ``non-geometric" structures dual to other (e.g. multipartite) patterns of entanglement) are woven into classical geometries?
	\end{itemize}

	These questions are related to the one posed in section 1.2.2 whose answer we are lacking, how in general to go beyond the classical Riemannian description of spacetime. 
			
				Of these questions, assuming that ER = EPR is true and probing the string scale properties of the ``non-geometric ER bridge" in search of a definition of it and an eventual self-consistent picture is probably the most tractable.

	In fact, the logical status is reminiscent of the strong form of the AdS/CFT conjecture. There, as one dials the coupling and number of DOF's in a CFT, a bulk dual description goes from a geometric to non-geometric one, and the latter has no meaning absent a independent formulation of strongly coupled quantum gravity. Rather it is taken to define strongly coupled quantum gravity. 
	 ER = EPR is a bulk-bulk non-supersymmetric \footnote{We can set up a wormhole for the BPS black hole, i.e. take the TFD state of the dual CFT's at $\beta = \infty$ which is some superposition of states on the moduli space. This is supersymmetric. But the ``classical wormhole" one gets from entangled BPS black holes is itself degenerate, it is infinitely long \cite{Leichenauer:2014nxa} suggesting the two sides can be interpreted as disconnected, so it appears a less useful case to study.} duality. It is not one in the sense of section 1.2.1. We are not inverting any dimensionless parameters. But (special cases of) it is a situation where as we tune a continuous parameter, we switch between one description that we know how to work with and one that we do not.
	
	What parameter? One can use the correspondence principle \cite{Susskind:1993ws, Horowitz:1996nw} to continuously interpolate from black hole microstates to weakly-coupled, perturbative string theory states. Correspondence transition can be thought of as a phase transition reminiscent of the confinement/deconfinement transition in QCD, with the black holes as bound states of strings \cite{Giveon:2005mi}. The weak form of ER = EPR says that entangled black holes (which are always very stringy/quantum objects regardless of how small the bulk coupling is otherwise) are dual to a semiclassical wormhole that we know how to work with if bulk gravity is weak. As we take the correspondence limit, a priori there is no reason to expect some analog of the ER bridge which is a feature of the ``confined" phase to survive in the ``deconfined" one, but this is claimed by the strong form of ER = EPR. On the other side of the transition the classicality of the wormhole interpretation is unclear, but the entangled black hole microstates become entangled perturbative string states which are in principle understood. 
	
	In various implementations of the correspondence principle, the string-black hole transition occurs as a function of different parameters.
		The original version \cite{Horowitz:1996nw} argued for a transition from the string to BH phase as $g_s$ is dialed from small to large, but that example is not under control; the authors only checked that the density of states of the black hole and perturbative string states were roughly of the same order at the transition point (when the BH radius equals $\ell_s$). Ref. \cite{Giveon:2005mi} contains an implementation in AdS$_3$/linear dilaton backgrounds where one dials the ratio $k^2 = R/\ell_s$ with $R$ the spacetime curvature. The theory is nonetheless perturbatively solvable because the worldsheet CFT is exactly known. 
		The black hole phase crosses over to the string one when the string length exceeds the AdS radius.	Very recently, \cite{Giveon:2015raa} pointed out another controlled example of a time-dependent black hole-string transition.

	

	
	

	An idea towards a first exploration of the properties of the non-geometric wormhole (assuming it exists) is to study maximally entangled AdS$_3$'s or linear dilaton theories, say in the thermofield state of their holographic dual, on the side of the correspondence transition without black holes. A similar suggestion was made in \cite{Balasubramanian:2014gla} to study entangled AdS's at temperatures below the Hawking-Page transition. This also would contain a ``non-geometric wormhole" by the ER = EPR conjecture. 
	However, to study features of the non-geometric wormhole we need to be able to compute at the string scale. 
	Here perhaps we can take advantage of exact worldsheet results (for the Euclidean, one-sided linear dilaton theory), following along the lines of recent scattering experiments \cite{Giveon:2015cma, Ben-Israel:2015mda} for Euclidean black holes, granted we can understand the analytic continuation to the Lorentzian, two-sided background. 




\subsection{Section summary}

Let us summarize in a page what one might mean by ``entanglement builds spacetime". 

There is the idea that position space entanglement in the CFT is the organizing principle explaining bulk geometry in AdS/CFT. The entanglement shadow is a problem for this idea. 

The more general idea is ``ER = EPR". It comes in two forms. The weak form says that entangled large black holes are connected in the interior by a non-traversable semiclassical wormhole. Why we believe that such a thing is true is because a Euclidean path integral argument constructs it fairly explicitly in AdS/CFT. 

The strong form replaces the black holes with more general bulk DOF's and the semiclassical wormhole with a ``non-geometric" one. It postulates that bulk entanglement builds (non-classical generalizations of) spacetime (that coalesces into classical spacetime when there are large amounts of particular entanglement patterns). The awkwardness of the phrasing reflects the ill-defined nature of the conjecture. When pulled back to the boundary in AdS/CFT, it suggests some form of boundary entanglement that need not be organized geometrically on the boundary (modulo ``subregion duality" crudely relating bulk and boundary locality), 
is the organizing principle underlying the bulk geometry. 

There is a hand-waving argument that the weak form implies the strong form by continuity, but we are missing a well-defined statement of the strong form or even a definition of the non-geometric wormhole. How to define a ``non-geometric wormhole" is related to ``how to define string geometry beyond Riemannian geometry." It is a hard string-scale problem.

 Towards clarifying the role of entanglement for emergence of spacetime, in our opinion the most promising directions are as follows. In the context of holography, (a) continue to study the problem of bulk reconstruction in all its forms 
 ; (b) understand precisely the subregion duality, that seems to work well around the vacuum while also going hand in hand with the limits of position-space entanglement in the CFT to describe the dual geometry, and thus appears to be a clue;
 %
 %
  and (c) define and study implications for bulk reconstruction of refined versions of entanglement such as the ``entwinement" \cite{Balasubramanian:2014sra}, whose claim to fame over other bulk probes is the absence of the shadow. In the context of working directly in the bulk, it is difficult to understand the strong form of the ER = EPR conjecture absent a definition of the ``non-geometric ER bridge". We suggested a naive approach is to assume the conjecture is true and then study the behavior of string-scale probes in backgrounds that are supposed to contain a non-geometric ER bridge, in hopes of identifying some of its properties. But perhaps the most fruitful way to approach this problem is indirectly, by continuing to develop our understanding of emergent geometry in string theory.

\section{Outline of the thesis}

In the rest of the thesis, I present two results.

In quantum systems, there are many (in)equalities that entanglement entropy can rigorously be shown to obey. For CFT's with an Einstein gravity dual, such equations can be translated to constraints on allowed bulk geometries using the Ryu-Takayanagi formula. I will show that two such inequalities, called positivity and monotonicity of the relative entropy translate to positivity conditions on the bulk matter energy density in any holographic dual, assuming bulk Einstein gravity. In classical general relativity, such energy conditions are usually postulated; here we derive them from first principles of the boundary theory. (See \cite{Lashkari:2014kda, Lashkari:2015hha} for related, independently-derived results).

Moreover, I will show that the Ryu-Takayanagi formula can be inverted for any state in the CFT to compute the local classical bulk stress-energy tensor at bulk points near the AdS boundary, in terms of entanglement across ball-shaped regions in the CFT. 

The results in the next chapter all come from \cite{Lin:2014hva}.

\chapter{Tomography From Entanglement}
This chapter is based on the paper \cite{Lin:2014hva} which was an equal collaboration with Matilde Marcolli, Hirosi Ooguri and Bogdan Stoica.  It should not be cited without citing that paper. 

In section 2.1, I review the entanglement inequalities that will be translated to the bulk, as well as what the constituents of those inequalities map to under the AdS/CFT dictionary. I state the assumptions made and range of validity of the results. In section 2.2 I review how one can use entanglement dynamics on the CFT boundary to derive the linearized Einstein equations around empty AdS in the bulk \cite{Lashkari:2013koa, Faulkner:2013ica}. In section 2.3 I demonstrate that an extension of techniques used there allows one to derive classical bulk energy conditions from entanglement inequalities in the boundary theory, and in section 2.4, I show that the local classical bulk stress-energy tensor at points near the boundary carn be expressed in terms of the boundary relative entropy. I conclude with a discussion in section 2.5.

Throughout this chapter, I use the following notation for the Ryu-Takayanagi formula: the entanglement entropy between a spatial domain $D$ of a CFT and its complement equals the area of the bulk minimal surface $\Sigma$ homologous to it, 
\be\label{rt}
S_{EE} = \min_{\partial D = \partial\Sigma}\frac{{\rm area}(\Sigma)}{4G_N}\,.
\ee
The volume enclosed by the Ryu-Takayanagi surface is denoted $V$.

\section{Setup} 

\subsection{Definitions in quantum information theory}

Relative entropy (see \eg~\cite{Blanco:2013joa}) is a measure of distinguishability between  quantum states in the same Hilbert space. The relative entropy of two density matrices $\rho_0$ and $\rho_1$ is defined as
\be \label{rele}
S(\rho_1 | \rho_0) = \tr(\rho_1 \log \rho_1) - \tr(\rho_1 \log \rho_0)\,.
\ee
Note that it is asymmetric in the arguments. It is positive, and increases with system size:
\begin{eqnarray} \label{positivity}
S(\rho_1|\rho_0) &\geq & 0\,,~~ \\
 S(\rho_1^W |\rho_0^W) &\geq & S(\rho_1^V|\rho_0^V), \qquad W \supseteq V\,.
\end{eqnarray}
The second property (2.4) is called monotonicity. 
When $\rho_0$ and $\rho_1$ are reduced density matrices on a spatial domain
$D$ for two states of a QFT, which is the case that we specialize to from here on, monotonicity implies that $S(\rho_1|\rho_0)$ increases with the size of $D$. That is, over a family of scalable domains with characteristic size $R$, 
\be\label{monotonicity}
\partial_R S(\rho_1|\rho_0) \geq 0\,.
\ee

Defining the modular Hamiltonian $H_{mod}$ of $\rho_0$ implicitly through
\be\label{hmod}
\rho_0 = \frac{e^{-H_{mod}}}{\tr(e^{-H_{mod}})}\,,
\ee
Eq. \eqref{positivity} is equivalent to 
\be\label{1stlaw}
S(\rho_1 |\rho_0) = \Delta \langle H_{mod}\rangle - \Delta S_{EE} \geq 0
\ee
where $\Delta\langle H_{mod}\rangle = \tr(\rho_1 H_{mod}) - \tr (\rho_0 H_{mod})$ is  the change in the expectation value of the operator $H_{mod}$ \eqref{hmod} and $\Delta S_{EE} = -\tr(\rho_1 \log \rho_1) + \tr(\rho_0 \log \rho_0)$ is the change in the entanglement entropy across $D$ as one goes between the states.

When the states under comparison are parametrically close, the positivity \eqref{1stlaw} is saturated to leading order \cite{Blanco:2013joa}:
\be \label{1stlawi}
S(\rho_0 + \delta\rho |\rho_0) = \delta \langle H_{mod}\rangle - \delta S_{EE}=0\,.
\ee
To see this, consider a reference state of the QFT characterized by $\rho_0$ and another, arbitrary state, $\rho_1$. We can construct a family of interpolating density matrices
\be\label{lambdaexp}
\rho(\lambda) = (1 - \lambda)\rho_0 + \lambda \rho_1
\ee
where $\lambda$ can be positive or negative. Because the relative entropy $S(\rho_0|\rho(\lambda))$ is positive for either sign of $\lambda$, the first derivative of this relative entropy with respect to $\lambda$ vanishes. This implies \eqref{1stlawi} to linear order in $\lambda$. 

Eq \eqref{1stlawi} is sometimes called the ``entanglement first law" for its resemblance to the first law of thermodynamics. Indeed, when $\rho_0$ is a thermal density matrix $\rho_0 = e^{-\beta H}/\tr(e^{-\beta H})$ with $\beta$ the inverse temperature, 
\eqref{1stlawi} reduces to $\Delta \langle H \rangle = T \Delta S$, an exact quantum version of the thermal first law. 

\subsection{The modular Hamiltonian}

In general, the modular Hamiltonian \eqref{hmod} associated to a given density matrix is a complicated operator that is not known exactly. \footnote{Although the leading order contribution to the modular Hamiltonian for the reduced density matrix of a spatial region in a QFT, expanded around the edge of the region, is always the Rindler Hamiltonian \cite{Bianchi:2012ev}.} There are a few simple cases where it is known. The most basic is when the density matrix is for a global thermal state of temperature $T$; then $H_{mod} = H/T$, where $H$ is the ordinary Hamiltonian.
Another is the reduced density matrix for the vacuum state of any QFT on the Rindler wedge. Since the Rindler wedge is thermal with respect to the boost generator, this case is isomorphic to the first one, with the modular Hamiltonian being proportional to the boost generator. 

Other cases are related to these by symmetry. 
The example that we'll use the most is when $\rho_0$ is the reduced density matrix on a ball of radius $R$ which we take to be centered at $\vec x_0 = 0$, in the vacuum state of a CFT  \cite{Casini:2011kv}. By the CHM map that was discussed in the previous chapter,
\be \label{defhmod}
H_{mod} = \pi\intop_D d^{d-1}x\frac{R^2-\left|\vec x \right|^2}{R}T_{tt}(x)\,,
\ee
where $T_{\mu\nu}$ is the energy-momentum tensor of the CFT.

\subsection{Assumptions}

Now I specialize to holographic CFT's. The starting point is a $d$-dimensional CFT whose vacuum state is dual to AdS$_{d+1}$ (suppressing the compact directions). I will assume the CFT has a semiclassical bulk dual where in particular the Ryu-Takayanagi formula holds around the vacuum.
To talk about the relative entropy \eqref{rele}, we have to compare this vacuum  state to another state in the Hilbert space. I take the latter to be an arbitrary excited state of the CFT. Einstein gravity probably breaks down somewhere in the holographic description of this state, but very near the boundary, the bulk still looks like AdS. I parametrize the near-boundary metric in the Fefferman-Graham form,
\be\label{fge}
g_{AdS} = \frac{\ell_{AdS}^2}{z^2}\lsb dz^2 + \lb\eta_{\mu\nu}+ h_{\mu\nu}\rb dx^\mu dx^\nu\rsb\,.
\ee
Spacetime indices $a, b, \dots$ run over $(z, t, x^i)$ while $\mu, \nu, \dots$ run over $(t, x^i)$ and $i \in 1, \dots, d-1$ are boundary spatial directions. 

\section{Review: Linearized Einstein equations from entanglement first law}

In this section, I review a result by van Raamsdonk and collaborators \cite{Lashkari:2013koa, Faulkner:2013ica} that the entanglement first law \eqref{1stlawi} for ball-shaped domains,
\be\label{ee1stlaw}
 \pi\intop_D d^{d-1}x\frac{R^2-\left|\vec x \right|^2}{R}\Delta \langle T_{tt}(x)\rangle = \Delta S_{EE}\,,
\ee
which can be viewed as an infinite set of constraint equations on states near the vacuum in a CFT, translates to the linearized Einstein equations about the AdS vacuum in the bulk. Mapping the entanglement first law to the bulk using the Ryu-Takayanagi formula gives one nonlocal equation for each bulk point, where the equation for each point comes from the entanglement first law for the Ryu-Takayanagi surface whose apex passes through that point. Our goal is to extract a local equation from this.

\subsection{$\Delta H_{mod}$ and $\Delta S_{EE}$ in holography}

We first have to explain how the semiclassical AdS/CFT dictionary maps the quantities appearing on both sides of \eqref{ee1stlaw} to the bulk.

The left-hand side of \eqref{ee1stlaw} is a weighted integral of the expectation value of the CFT stress tensor $\langle T_{tt}\rangle$. In the vacuum state of the CFT, this is known. In the excited state, $\langle T_{tt} \rangle$ can be expressed as a function of bulk fields by using holographic renormalization \cite{Balasubramanian:1999re, Skenderis:2002wp}, where fixing $\langle T_{tt}\rangle$ in the vacuum sets the holographic renormalization scheme.

Alternatively, \cite{Faulkner:2013ica} pointed out a short-cut to obtain the holographic dictionary entry for the boundary stress tensor of the CFT, by exploiting the fact that the relative entropy in the CFT vanishes in the limit that the entangling domain shrinks to zero size. The trick is to take $R \rightarrow 0$ in \eqref{ee1stlaw} and map the right-hand side to the bulk using the RT formula, keeping the left-hand side fixed as the definition of $\Delta \langle T_{tt}\rangle$ . 

As long as the bulk matter fields contributing to $\langle T_{\mu\nu} \rangle$ are dual to operators 
with scaling dimension $\delta > d/2$, both methods give
\be
\label{hmode}
\Delta \langle H_{mod}\rangle = \lim_{z \rightarrow 0} \frac{d\ell^{d-1}_{AdS}}{16 G_N} \intop_D d^{d-1}x \frac{R^2-|\vec x |^2}{R} z^{-d} \eta^{ij} h_{ij} \,. \qquad
\ee
As usual, operators with dimension $\delta \leq d/2$ require a more careful treatment \cite{Klebanov:1999tb}.

To translate $\Delta S_{EE}$ on the right-hand side of \eqref{ee1stlaw} to the bulk, we use the Ryu-Takayanagi formula. On a constant-time slice of pure AdS, the codimension-2 bulk extremal surface $\Sigma$
ending on a boundary sphere of radius $R$ is the half-sphere
\be\label{minads}
z_0(r) = \sqrt{R^2-r^2}\,.
\ee

The EE of the entangling disk of radius $R$ in the CFT vacuum is equal to the area functional of pure AdS evaluated on the surface \eqref{minads}.
Suppose we perturb the bulk metric away from pure AdS by $h_{ab}$ which is parametrically small. Because the original surface was extremal, the leading variation in the holographic EE comes from evaluating $z_0(r)$ \eqref{minads} on the perturbed area functional. One finds \cite{Lashkari:2013koa}
\be \label{seed}
\Delta S_{EE} = \frac{\ell^{d-1}_{AdS}
}{8 G_N R}\intop_{\Sigma} d^{d-1}x  (R^2 \eta^{ij} - x^ix^j )z^{-d}h_{ij}. \quad
\ee
At order $h^2$, one must account for corrections to the shape of the Ryu-Takayanagi
surface; see \eg\ \cite{Blanco:2013joa}.

\subsection{Linearized Einstein equations}

We have now translated \eqref{ee1stlaw} to one nonlocal bulk constraint on the linearized metric perturbation $h_{ab}$ for each ball $B$ in the boundary CFT. The authors of \cite{Faulkner:2013ica} noticed that one can write $\Delta H_{mod}$ \eqref{hmode} and $\Delta S_{EE}$ \eqref{seed} as integrals over the entangling disk $D$ on the boundary and the extremal surface $\Sigma$ in the bulk, respectively, of a local ($d-1$)-form $\pmb\chi$ that is a functional of $h_{ab}$:
\be
\label{faulkner}
\intop_D \pmb\chi = \Delta\langle  H_{mod}\rangle , \qquad \intop_{\Sigma}\pmb\chi = \Delta S_{EE}\,.
\ee 	
%
Moreover, the exterior derivative of this $\pmb\chi$ is given by
\be\label{dchi}
{\rm d}{\pmb\chi} = 2\xi^t  E^g_{tt}[h] g^{tt} \sqrt{g_V} dz\wedge dx^{i_1}\dots\wedge dx^{i_{d-1}},
\ee
where $\sqrt{g_V}$ is the natural volume form on $V$ induced from the bulk spacetime metric, and 
\be\label{kv}
\xi =  \frac{\pi}{R}\left\{ [R^2\!-\!z^2\!-\!(t-t_0)^2\! -\! x^2]\partial_t \! - 2(t  -  t_0)[z\partial_z  +  x^i\partial_i] \right\} 
\ee
is the Killing vector associated with $\Sigma$ \eqref{minads}, which is a Killing horizon in pure AdS. The linear gravitational equations of motion in vacuum are
expressed as  $E^g_{ab}[h]=0$.

By the Stokes theorem, the relative entropy is given by
\be\label{eqone}
S(\rho_1|\rho_0) = \Delta\langle H_{mod}\rangle - \Delta S_{EE} = \intop_\Sigma {\pmb\chi} - \intop_D {\pmb \chi}
= \intop_V {\rm d}\pmb\chi\,. 
\ee
Considering \eqref{eqone} for every disk on a spatial slice at fixed time $t=0$, the entanglement first law $S(\rho_1|\rho_0)=0$ can be shown to be equivalent to  $ E^g_{tt}[h] = 0$. 
Considering it for Lorentz-boosted frames gives vanishing of the other boundary components, $ E^g_{\mu\nu}[h] = 0$.
Finally, appealing to the constraint equations of the initial-value formulation of gravity gives the vanishing of the remaining components of the linearized Einstein tensor that carry $z$ indices. This completes the proof that for holographic theories where the Ryu-Takayanagi formula holds, the linearized Einstein equations around vacuum AdS are dual to the entanglement first law around the vacuum in the CFT.

Of course, the entire argument as presented, rests on the existence of the form ${\pmb \chi}$ with the advertised properties. This object was constructed by Iyer and Wald \cite{Wald:1993nt, Iyer:1994ys} to prove the converse for black hole horizons: that if one perturbs a black hole solution in accordance with the linearized Einstein equations, the change in the black hole horizon area equals the change of its gravitational energy. The applicability of their work to our situation relies on the fact that the Ryu-Takayanagi surface for a ball-shaped region in a CFT vacuum is a cross-section of the AdS-Rindler wedge, which (as discussed in section 1.5.4) is a massless hyperbolic black hole in empty AdS \cite{Emparan:1999gf}. 

By repeating the derivation with the first $1/N$ correction to the Ryu-Takayanagi formula \cite{Faulkner:2013ana}, Swingle and van Raamsdonk showed \cite{Swingle:2014uza}  that the entanglement first law \eqref{1stlawi} implies  the linearized bulk Einstein equations sourced by the difference in the expectation value of the bulk stress-energy tensor in the quantum state of bulk fields relative to their vacuum state, $\delta \langle t_{ab}\rangle$. (The key technical step they used is that one can compute the bulk modular Hamiltonian of the AdS-Rindler wedge as well.)
	They argued that assuming the source of the linearized Einstein equation is a local operator, one can lift $\delta\langle t_{ab}\rangle$ to the bulk operator $t_{ab}$ in their argument, and subsequently (by consistency of adding the coupling term $t_{ab}h^{ab}$ to the Lagrangian, assuming the bulk theory is described by a local action) the linear Einstein equations to the nonlinear ones.
	In the discussion, I will comment on prospects for deriving the nonlinear Einstein equations without assuming bulk locality.

In the rest of the chapter, {\it I will assume that the linearized Einstein equations coupled to the bulk stress tensor hold in the bulk}, and derive energy conditions from them.


\section{Bulk energy conditions from relative entropy}

On trying to apply the previous line of reasoning to general states in the CFT, we immediately encounter a technical problem: by going to order $\lambda^2$ in the expansion \eqref{lambdaexp}, we would have to account for the quadratic fluctuation of the bulk metric sourced by the boundary stress tensor, which will change the shape of the Ryu-Takayanagi minimal surface away from the AdS-Rindler cross-section, that was crucial for the definition of the form ${\pmb\chi}$  in the previous section. 

 To continue to apply the same formalism, we will take a different limit, where we consider arbitrary states in the boundary CFT but restrict to entangling domains whose radii $R$ are small compared to the typical energy scale $\EE \approx \langle T_{\mu\nu}\rangle^\frac{1}{d}$ of the state measured by the boundary stress tensor $T_{\mu\nu}$, {\it i.e.} $ \EE R \ll 1$. So the Ryu-Takayanagi surface only penetrates the asymptotically AdS region of the bulk.

In the interior of the Ryu-Takayanagi surface for such entangling domains, we can evaluate the $(d-1)$-form ${\pmb \chi}$ of \cite{Faulkner:2013ica} on the bulk metric fluctuation $h_{ab}$ 
of the dual to an arbitrary excited state of a CFT. As the deviation of the bulk metric in the enclosed volume $V$ from pure AdS is parametrically small, all results of the above discussion carry over. ${\pmb \chi}$ integrated over the entangling ball on the boundary gives $\Delta H_{mod}$, ${\pmb \chi}$ integrated over the semi-circular RT surface gives $\Delta S_{EE}$, and $d{\pmb\chi}$ satisfies \eqref{dchi}.  
The Stokes theorem now reads
\be\label{eqtwo}
S(\rho_1|\rho_0) = \Delta\langle H_{mod}\rangle - \Delta S_{EE} 
= \intop_\Sigma {\pmb\chi} - \intop_D {\pmb \chi} = \intop_V {\rm d}\pmb\chi\, \leq 0. 
\ee
At the same time,
$E^g_{ab}[h]$ in  \eqref{dchi} should be  evaluated on  
the $h_{ab}$ which is reconstructed from CFT data at non-linear level and is not identically zero. 
Rather, the linearized Einstein tensor couples to bulk matter in the form of the bulk stress tensor,
\be\label{efe}
E^g_{ab}[h_{ab}] = 8\pi G_N t_{ab}\,.
\ee

Using \eqref{dchi} we find that the relative entropy is expressed as the
integral of the local bulk energy density $\varepsilon  = - t^{t}_{~t}$, 
\be\label{wec}
S(\rho_1|\rho_0) = 8\pi^2 G_N\intop_V \frac{R^2-(z^2+x^2)}{R} \varepsilon \sqrt{g_V} \, .
\ee
As one pretty interesting added note, Casini \cite{casini} pointed out to us that the expression on the right-hand side of \eqref{wec} is related directly to the modular Hamiltonian of the AdS-Rindler wedge of the bulk.


Taking one derivative with respect to $R$, we find
\be \label{wectwo}
\partial_R S(\rho_1 | \rho_0) = 8\pi^2 G_N\intop_V\left(1 + \frac{z^2+x^2}{R^2} \right)\varepsilon \sqrt{g_V}
 \,.
\ee
Though the derivative also generates an integral over the Ryu-Takayanagi surface $\Sigma$, it vanishes because $\xi^t$ vanishes on the surface. 

Recall that positivity $S \geq 0$ and monotonicity $\partial_R S \geq 0$ are universal properties of the 
relative entropy \eqref{positivity}. Comparing to \eqref{wec}, \eqref{wectwo} we find that, in the gravitational dual, they are translated to positivity of the 
integrals of the bulk energy density $\varepsilon$ weighted by 
$(R^2 \pm (z^2 + x^2))\sqrt{g_V}$ ($\geq 0$
in $V$). These are derivations of integrated weak energy conditions in the bulk from boundary principles.

One more derivative relates the relative entropy to  
the integral of the energy density on the boundary $\Sigma$ of $V$,
\be\label{mon}
\left(\partial^2_R + R^{-1} \partial_R - R^{-2} \right) S(\rho_1|\rho_0) = 16\pi^2 G_N \intop_\Sigma \varepsilon\sqrt{g_\Sigma} \,, \qquad
\ee
where $\sqrt{g_\Sigma}$ is the volume form on the Ryu-Takayanagi surface. I will now show that 
\eqref{mon} can be inverted using the Radon transform 
to express the bulk stress tensor point-by-point in the near-AdS region using the entanglement information of the CFT.

\section{Local bulk stress tensor from relative entropy} We found that $\partial_R S(\rho_1|\rho_0)$ is given by the integral of the energy density $\varepsilon$
over the region $V$ inside the Ryu-Takayanagi surface. We can invert this relation to compute 
$\varepsilon$ point-by-point in the bulk by using the relative entropy $S(\rho_1|\rho_0)$.

To show this, note that 
\be\label{rd}
\left( \partial_R + R^{-1} \right) S(\rho_1|\rho_0) = 16\pi^2 G_N\intop_V\varepsilon\sqrt{g_V}
\ee
so differentiating again,
\be\label{radon}
\left(\partial^2_R + R^{-1} \partial_R - R^{-2} \right) S(\rho_1|\rho_0) = 16\pi^2 G_N \intop_\Sigma \varepsilon\sqrt{g_\Sigma} \qquad
\ee
where $\sqrt{g_\Sigma}$ is the natural volume form on the Ryu-Takayanagi surface $\Sigma$
induced from the bulk
spacetime metric. The right-hand side is still non-negative if we assume the positivity of the bulk energy density. Thus, 
\be
\left(\partial^2_R + R^{-1} \partial_R - R^{-2} \right) S(\rho_1|\rho_0) \geq 0.
\ee

Here the bulk geometry is empty AdS, and its space-like section is $d$-dimensional 
hyperbolic space. The surface $\Sigma$ is then totally geodesic.
In this case, the integral \eqref{radon} is the Radon transform and its inverse is known. 
The Radon transform can be thought of as a souped-up Fourier transform in hyperbolic space.
For a smooth function $f$ on $d$-dimensional hyperbolic space, the Radon transform ${\cal R}f(\Sigma)$ is the integral of $f$ over an $n$-dimensional 
geodesically complete submanifold $\Sigma$ with $n< d$. The dual Radon transform ${\cal R}^* {\cal R}f$ gives back 
a function on the original hyperbolic space in the following way: pick a point in the hyperbolic space, consider
all geodesically complete submanifolds passing through the point, and integrate ${\cal R}f$ over such
submanifolds.
It was shown by Helgason \cite{Helgason} that if $d$ is odd, $f$ is obtained by applying an appropriate differential operator on 
${\cal R}^* {\cal R}f$. We are interested in the case $n = d-1$ for which
\be
  f = \lsb (-4)^{(d-1)/2}\pi^{d/2-1}\Gamma(d/2) \rsb^{-1} Q({\bf \Delta}) {\cal R}^* {\cal R}f\,, \ 
\ee
where $Q({\bf \Delta}) $ is constructed from the Laplace-Beltrami operator ${\bf \Delta}$ on the hyperbolic space as
\be
Q({\bf \Delta}) = \lsb{\bf \Delta} + 1\cdot (d-2)\rsb\lsb{\bf \Delta}+2\cdot (d-3)\rsb 
\times \cdots \times \lsb{\bf \Delta} + (d-2)\cdot 1\rsb \, .
\ee

Applying this to \eqref{radon}, we find 
\be\begin{split}
\label{Helgason}
\varepsilon = \lsb (-4)^{(d+3)/2} \pi^{d/2+1}
\Gamma(d/2) G_N \rsb^{-1} \times \\ 
 Q({\bf \Delta}) {\cal R}^* \left(\partial^2_R + 
R^{-1} \partial_R - R^{-2} \right) S(\rho_1|\rho_0) \,, 
\end{split}
\ee
when $d$ is odd. There exists a similar formula for $d$ even~\cite{Rubin}.

\begin{figure}[h]
\centering
\includegraphics[width=0.9\textwidth]{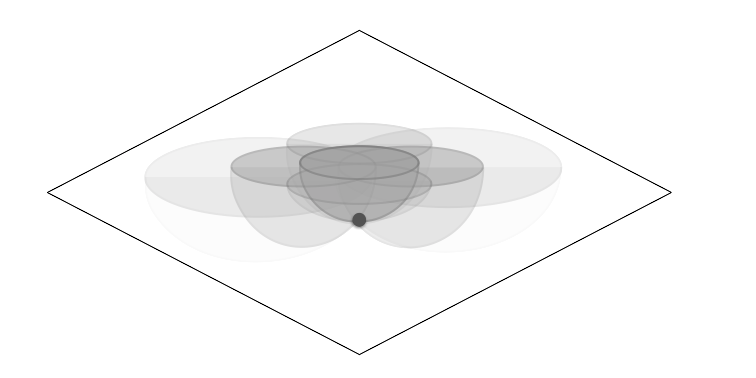} 
    \caption{The bulk energy density at a point near the boundary of AdS can be approximately extracted from the entanglement entropies across ball-shaped domains of the boundary, whose Ryu-Takayanagi surfaces pass through the point.}
\end{figure}

Note that even if we are evaluating $\varepsilon$ at $(z, t, x)$ in the near-AdS region, there are 
totally geodesic surfaces that pass through this point and go deep into the bulk, where the geometry can
depart significantly from AdS. However, contributions from these surfaces are negligible  
when $\EE z \ll 1$, where $\EE$ is the
typical energy scale of the CFT state. In this case, we can choose another $z_0$ so that $z \ll z_0$ and
the geometry under $z_0$ is still approximately AdS. Since most totally geodesic surfaces passing 
through $(z,t,x)$ stay under $z_0$,  an integral over such surfaces is well-approximated by 
the inverse Radon transform in the hyperbolic space. 

The energy density is the time-time component of the stress-energy tensor $t_{ab}$. By computing the relative entropy in other Lorentz frames, we can also derive components $t_{\mu\nu}$ along the boundary. Finally, we can use the conservation law, $\nabla^a t_{ab}=0$, to obtain the remaining components, $t_{z\mu}, t_{\mu\nu}$. So
we can use the entanglement data on the boundary to reconstruct all components of the bulk stress 
tensor. 

We emphasize that the energy density thus obtained is classical. It is not the expectation value of the bulk quantum stress tensor, and hence not an example of a bulk local operator reconstruction in holography.

\section{Discussion}

We showed two main results, which are a classical energy condition in the near-AdS region of the bulk from an entropy inequality in the dual CFT, and determining the local classical energy density at points in the near-AdS region from nonlocal entanglement data in the CFT. In principle, we could attempt to push either result further into the bulk interior (see \cite{Lashkari:2015hha} for some progress in this direction). 

I conclude with brief comments about direct follow-up ideas.
\begin{itemize}
\item [$\diamond$] {\it Can we promote the energy condition to be a quantum one in the bulk?} 
\end{itemize}
No. As explained in \cite{Swingle:2014uza}, the expectation value of the bulk stress tensor from the state of bulk quantum fields can be consistently included in the above derivation by including the $1/N$ correction to the Ryu-Takayanagi formula. If we kept this correction in equation (2.21), we would find that the RHS contains actually $8\pi G_N (t_{ab} - \langle t_{ab}\rangle)$ with the latter the expectation value of the bulk stress tensor relative to the vacuum. I.e. if we assumed the linearized Einstein equations coupled to the full classical + quantum stress tensor at the end of section 2.2 (instead of assuming just the linearized Einstein equations coupled to the classical stress tensor), but also consistently included the $1/N$ correction to the Ryu-Takayanagi formula which contributes at the same order, the quantum part of the energy density would cancel out, leaving the conclusion unchanged that we derive (2.22), (2.23) with $\varepsilon$ the classical energy density. 

\begin{itemize}
\item[$\diamond$] {\it What is the geometric meaning of the CFT relative entropy w.r.t. the vacuum state of the CFT, beyond the limit described above?} 
\end{itemize}
	We can make a few simple predictions. It will be a classical bulk inequality that will be completely geometric. Using the Einstein equations, we can convert some functionals of the bulk metric to functionals of the matter stress-energy tensor, and indeed it has to reduce to such in the limit discussed above. On the other hand, the AdS black hole is a counterexample to the possibility that the gravity dual of positivity of relative entropy is always a positivity condition on the matter energy density. So the gravity dual of the relative entropy inequalities should be positivity conditions on a classical linear combination of matter and gravitational energy, which indeed has been borne out so far by the second-order study of \cite{Lashkari:2015hha}.
	
	
	\begin{itemize}
\item[$\diamond$] {\it Can the nonlinear Einstein equations be derived from position-space entanglement in the CFT?} 
\end{itemize}

Not for the reasons explained in Section 1.5.3.
 To be a bit more precise, we can ask two separate questions. Can the RT formula be used to show that the classical bulk stress tensor sources the Einstein equations around the AdS vacuum? No, because the RT formula is a static one that only ever sees geometry  -- namely the metric -- in the bulk. One has to put in preexisting knowledge about dynamics (as indeed we did at the end of section 2.2) to know that the classical bulk stress tensor is responsible for the backreaction.
	How about the nonlinear Einstein equations in vacuum? This would appear to be ruled out by the entanglement shadow (e.g. in black hole backgrounds).
	
As advocated previously, if information-theoretic ideas in the CFT are indeed key to decoding the hologram, the fundamental probe must not be the position-space entanglement, but a different, related quantity. Indeed it is even logically possible that the entanglement first law $\delta S_{EE} = \delta \langle H_{mod}\rangle$ around excited states can be translated to the Einstein equations if the modular Hamiltonians of subregions in high-energy CFT states are sufficiently complicated operators to probe the full bulk geometry, but then they, not the position-space entanglement itself as codified in the Ryu-Takayanagi formula, are the heroes of the story. 

\makebibliography

\end{document}